\documentclass[12pt]{iopart}

\usepackage{color}
\usepackage{graphicx}
\usepackage{listings}
\usepackage{subfigure}
\usepackage{color}
\usepackage{float}
\usepackage[T1]{fontenc}

\begin{document}

\title{Short term fluctuations of wind and solar power systems}

\author{M.  Anvari $^{1}$, G.  Lohmann $^{1}$, M. W{\"a}chter $^{1}$, P.  Milan $^{1}$, E. Lorenz $^{2}$, D. Heinemann $^{1}$, M. Reza Rahimi Tabar $^{1,3}$, Joachim Peinke $^{1}$,  }
\address{$^{1}$ Institute of Physics and ForWind, Carl von Ossietzky University, 26111 Oldenburg, Germany}
\address{ $^{2}$ Institute of Physics, Carl von Ossietzky University, 26111 Oldenburg, Germany}
\address{ $^{3}$ Department of Physics, Sharif University of Technology, Tehran 11155-9161, Iran}

\ead{Correspondence and requests for
materials should be addressed to J. P.,  peinke@uni-oldenburg.de  }
\vspace{10pt}

\begin{abstract}

Wind and solar power are known to be highly influenced by weather events and may ramp up or down abruptly. Such events in the power production influence not only the availability of energy, but also the stability of the entire power grid.
By analysing significant amounts of data from several regions around the world with resolutions of seconds to minutes, we provide strong evidence that renewable wind and solar sources exhibit multiple types of variability and nonlinearity in the time scale of {\it seconds} and characterise their stochastic properties.
In contrast to previous findings, we show that only the jumpy characteristic of renewable sources decreases when increasing the spatial size over which the renewable energies are harvested. Otherwise, the strong non-Gaussian, intermittent behaviour in the cumulative power of the total field survives even for a country-wide distribution of the systems. 
The strong fluctuating behaviour of renewable wind and solar sources can be well characterised by Kolmogorov-like power spectra and $q-$exponential probability density functions. Using the estimated potential shape of power time series, we quantify the jumpy or diffusive dynamic of the power. Finally we propose a time delayed feedback technique as a control
algorithm to suppress the observed short term non-Gaussian statistics in spatially
strong correlated and intermittent renewable sources.
\end{abstract}

\section{Introduction}

The renewable energy sources and their share in electricity production have increased constantly, mainly driven by energy policies, markets and environmental issues.  Among the renewable energy sources the use of wind power and photovoltaics (PV) has a priority. For instance in the European Union, these renewable energies shall account for about 20 \% of the gross final energy consumption by 2020 and 60 \% by 2050 \cite{1}. These renewable sources are commonly known to be highly intermittent, i.e.\ they are highly fluctuating on many different time scales, see \cite{2,2-2} and references therein. Therefore, one of the most important future challenges for the stability of a desired supply grid, based on renewable energies, will be  control and suppressing of these fluctuations. 
  
In traditional power plants, the inertia of fast rotating generators is utilised as an automatic power reserve. 
This is done simply by speeding up or slowing down the rotating masses, keeping the grid frequency within a narrow range around the nominal frequency. In the ENTSO-E \footnote{The ENTSO-E (European Network of Transmission System Operators for Electricity) is an association of European transmission system operators which covers virtually all of Europe.} grid, the value of the nominal frequency is $50$ Hz and the tolerated deviation from this value is $\pm 10$ mHz \cite{2-3}. 
Restoring the grid frequency to the nominal frequency, in current practice, is provided by traditional frequency control, which has three categories: primary, secondary and tertiary frequency control, cf.\ \cite{3}. 
The primary frequency control is provided within a few seconds after the occurrence of a frequency deviation. 
It provides extra power for stabilising the system frequency (but not restoring it to the nominal frequency $f_0$) \cite{3-1}. 
The secondary frequency control acts after approximately 30~s and restores both the grid frequency from its residual deviation and the corresponding tie-line power exchanges with other control zones to the set-point values. 
Tertiary frequency control manually adapts power generation and load set-points and controls the grid operation beyond the initial 15 minute time-frame after a fault event has occurred. 

In the background of replacing the successively controllable conventional power plants by intermittent renewable power systems, there are several recent works studying the grid stability under these new constraints \cite{4,5,5-1}. One practical approach is that synchronous machines of old power plants are still connected to the grid and providing the reactive power and inertia \cite{5-2}.
It has also been a practical topic to study how the stability of the power grid can be kept in the lower rotational inertia case (because of high penetration of renewable sources) using some faster control reserves \cite{6-1,6-2}. One possible option is to use battery storage providing primary control reserve, see e.g. \cite{6-3} for a very recent study on this topic.

Based on different aforementioned control techniques, one has to break up the grid stability consideration into different time scales of the fluctuating renewable sources. The most recent studies consider the fluctuations in wind and solar powers in 15 or 60 minutes and investigate the effects of these fluctuations in power system \cite{6-4,6-5} and the trading on the electricity market \cite{6-5,6-6,6-7}. However, up to now, little work has been done in connection with disentangling the time dependency of these fluctuations. This is the topic that we address in this paper and in particular we focus on short time scales. Indeed, we believe that understanding the renewable energy characteristics in short time scales will be an important additional aspect to design the efficient control systems in future power grids.

Generally, the short time fluctuations have been less investigated, as on the one hand it is hard to get the high-frequency power data (such as 1~Hz data), and on the other hand it is commonly assumed that the fast fluctuations average out geographically. Further for supply systems with big shares of  traditional power units  the primary and secondary reserve guarantee an easy automatic control. The situation of a power system with high shares of wind and solar energies is different, as for modern wind turbines the transfer of wind power to the supply grid is based on an AC/DC-DC/AC rectifier - inverter technique adapted the wind power to the supply grid conditions with 50/60 Hz \cite{4}. By this technique the inertia of the rotating part of a wind turbine is decoupled from the grid. Also PV systems do not automatically provide inertial response.
  
A future supply grid with low rotational inertia will have implications for operational instabilities of power systems \cite{7}. 
For instance, in Ireland's power grid, currently the share of renewables is strictly limited to \%50, because of the inertia problem \cite{7-1}. 
The complexity of future power grids with increasing shares of renewable sources requires a precise characterisation and understanding of the short term fluctuations of wind and solar installations in the time range of seconds. 
On this basis, new solutions can be worked out to suppress the undesired but natural fluctuations in more most efficient way.

In this contribution we will present results of time series analysis of a unique data set for power output from different solar and wind systems in several regions around the world with resolutions of seconds to minutes. 
The data set is ranging from power output of single power systems to the country wide power production. 
The data analysis is based on two approaches. 
On the one hand the characterisation of stochastic properties of power in different short time scales is performed using power and irradiance increments $X_{\tau} := X(t+\tau) - X(t)$.
From these we study how likely fluctuations of certain amounts will occur, for example $50\%$ of the rated power will emerge in a time lag $\tau$ in the order of a few seconds.
On the other hand the increment statistics are complemented by studying the temporal evolution of the power dynamics, as dynamical properties are not grasped completely by the statistical two-point quantities $X_{\tau}$.
Both methods will give new insights into the properties of the power fluctuations with respect to time scales and geographical averaging. Besides these new results, we also include some already published results about the characteristics in the short time fluctuations to complete the discussion of power dynamics.

This paper is organised as follows. 
In Sec. 2, we describe the analysed big data sets for wind power, solar power and solar irradiance data. 
In Sec. 3, we provide strong quantitative evidence that both wind and solar energy resources exhibit short time nonlinear variability which typically occurs at time scales of a few seconds and show that the intermittency and strong non-Gaussian behaviour in cumulative power of the total field still survives in both cases, even for a country-wide installation. 
In Sec. 4, using the potential shape of power time series, we find that depending on the spatial size over which the renewable energies are harvested, there is a critical phase transition of the stochasticity from jumpy, i.e. on-off type, to a persistent stochastic process. Also we used the potential analysis to detect the tipping point of this transition. As a conclusion of our data analysis, we propose in Sec. 5 a time-delayed feedback method for suppressing the short term extreme events of power output of wind farms and solar fields. In the new presented method we show that saving a portion of power output of a single renewable source, and injecting it after a delay of about $2-5$ seconds, will have noticeable impact on the short time intermittency. The paper is summarised in Sec. 6 and a resulting picture of high frequency power dynamics is presented.

\section{Description of high frequency data sets of wind power and solar irradiance} 

The paper is based on a large set of measurements of high-frequency data for renewable wind power, solar power and solar irradiance which are selected from different countries around the world (see Table 1). The sampling rates range from $0.001$Hz to $1$Hz. The data sets include wind and solar power and irradiance time series from wind farms and solar power plants with different sizes, which enables us to study the changes in their statistical properties as a function of the field size. 

The wind data were obtained from: 
\begin{itemize}
\item{W1-} wpd windmanager GmbH, Bremen which includes 12 turbines and spreads over a rectangular area of roughly $ 4\times 4$~km$^2$ \cite{2}, a subset of these data is available under \cite{datawind}.

\item{W2-} Tennet recording the whole wind energy production of Germany (here, the date between 2007 and 2012 has been used) \cite{data1}. 
\item{W3-} Eirgrid recording the whole wind production of Ireland (here, the date between 2007 and 2012 has been used) \cite{data2}.
\end{itemize}

The solar data were recorded from: 
\begin{itemize}
\item{S1-} An observational network on a platform roof of the University
of Oldenburg, Germany (53.152$^{\circ}$ N, 8.164$^{\circ}$ E). It
consists of up to $16$ small ($0.242 \times 0.556$~m$^2$ each)
photovoltaic (PV) modules spanning an area of about $250 \times
250$~m$^2$ and was used by and presented in \cite{dataS1}.  A subset of these data (clearsky index recorded by 11 sensors
in June 1993) is available under  \cite{datawind}.

\item{S2-} The United States' National Renewable Energy Laboratory (NREL) which performed a one-year measurement campaign at Kalaeloa Airport ($21.312^{\circ}$ N, -158.084$^{\circ}$ W), Hawaii, USA, from March 2010 until March 2011 using 19 LI-COR LI-200 pyranometers to measure global solar irradiance on horizontal and inclined surfaces \cite{dataS2}. Two of the instruments were tilted by 45 degrees, while the
other 17 were horizontally mounted and scattered across an area of about $750 \times 750$~m$^2$. The data is available from \cite{dataS2-2}. 
\item{S3 and S4-} The Baseline Surface Radiation Network (BSRN) where solar and atmospheric radiation are measured with instruments of the highest available accuracy and with high temporal resolution. Multi-year time series of global horizontal irradiance were available for one station (S3) in northern Spain recording data between July 2009 and February 2013, and one station \cite{dataS3} (S4) in Algeria (Sahara) recording data between March 2000 and December 2013. The station in Spain is situated in an urban environment in a mountain valley ($42.816^{\circ}$ N, $-1.601^{\circ}$ W), while the station in Algeria is surrounded by rock and desert ($22.790^{\circ}$ N, $5.529^{\circ}$ E) \cite{dataS3}. 
\item{S5-} Fraunhofer Institut f{\"u}r Solare Energiesysteme (ISE) recording the whole solar energy production of Germany in 2012.
\end{itemize}

\begin{table} 
\begin{center}

\caption{Data description}\label{Table1}\vspace{0.5cm}
\scalebox{0.8}{
\begin{tabular}{cccccc}
\hline\hline
 data set & rated power & data points & measurement duration & frequency   \\
  \hline 
 W1:  wind farm (12 turbines) & $\sim$ 25 MW & $15.3 \times 10^6$ & $\sim$ 8 months & 1 Hz   \\ 
   \hline
W2:  wind farm Germany   & $\sim$ 30 GW &  $  \sim 2 \times 10^5$ & $\sim$ 6 years & 1/15 $min^{-1}$   \\ 
\hline
W3 : wind farm Ireland   & $\sim$ 1000 MW &  $ \sim 10^6$ & $\sim$ 10 years & 1/15 $min^{-1}$   \\ 
 \hline
S1:    solar irradiance, Germany (Oldenburg) & -- &  $12 \times 10^6$ &  $\sim$16 months & 1 Hz  \\ 
  \hline
S2:  solar irradiance, Hawaii & -- &  $ 14 \times 10^6$ &  $\sim$12 months & 1 Hz  \\ 
  \hline
S3: solar irradiance, Spain & -- &  $ 1.3 \times 10^6$ &  $\sim$31 months & 1/60 Hz \\ 
\hline
S4: solar irradiance, Sahara & -- &  $ 3.7 \times 10^6$ &  $\sim$86 months & 1/60  Hz  \\ 
 \hline
S5:  solar field Germany   & $\sim$ 30 GW &  $  \sim 17000$ & $\sim$ 1 year & 1/15 $min^{-1}$  \\ 
\hline
\end{tabular}
	}
\end{center}
\end{table}

For the analysis of the recorded data sets we first scale these time series to have dimensionless data for drawing a comparison between the results. Therefore, we calculate the scaled wind power $P(t)/ P_r$, where $P_r$ is the rated power and the clear sky index $Z=G(t)/G_{clear sky}$, where $G(t)$ and $G_{clear sky}$ are the measured solar irradiance and its theoretical prediction under clear sky at a given latitude and longitude, respectively. We used the model presented in \cite{68} to compute the clear-sky index time series which needs to include parameters of atmospheric conditions, such as air composition and turbidity \cite{68}. The clear sky index has positive values and its maximum is around unity. 
\section{Intermittency: non-Gaussian behaviour of wind and solar increments statistics} 

In this section we focus on the characterisation of short time power fluctuations. 
We use a two-points statistics analysis based on increment statistics in lag $\tau$, i.e. $X_{\tau} := X(t+\tau) - X(t)$. 
The increment  $X_{\tau}$ may have positive and negative values corresponding to the ramp-up and ramp-down events as seen from the present state $X(t)$. The increment analysis can be done in two different ways. One may investigate the $\tau$- dependence of the increment moments, which is called the structure functions $S_n(\tau) := \langle \Delta X_{\tau} ^n \rangle $ \cite{13}. Alternatively, one may analyse the $\tau$- dependence of the probability density functions (PDF) $P(X_{\tau}, \tau)$, for which we use the short notation $P(X_{\tau})$. Note that the second order structure function $S_2(\tau) = \langle \Delta X_{\tau} ^2 \rangle $ is related to the autocorrelation $\langle X(t+\tau) \cdot X(t) \rangle$, which in turn is directly related to the power spectrum by a Fourier transform, after the Wiener-Khinchin theorem. This fact makes clear that the often used power spectra only characterise the $\tau$-dependence of the width or standard deviations  $\sigma_{\tau}^2 = \langle \Delta X_{\tau} ^2 \rangle $ of the PDFs $P(X_{\tau})$. A remarkable feature of the PDFs $P(X_{\tau})$ is that they show for many systems, in particular for turbulence-like systems (and for small values of $\tau$) pronounced deviations from Gaussianity. If the PDFs are heavy tailed with high probabilities of extreme events, we define this as intermittency, following the common notion for turbulence \cite{8}. This can also be quantified by higher order structure functions \cite{8,9,9b,10,11,12}. Consequently, we analyse here the wind and solar data sets with respect to the power spectra and the increment PDFs mainly for the normalised data sets, i. e. $X_{\tau} / \sigma_{\tau}$, where $\sigma_{\tau}$ is the standard deviation of $X_{\tau}$.

\begin{figure}
\centering
\includegraphics[width=0.3\hsize]{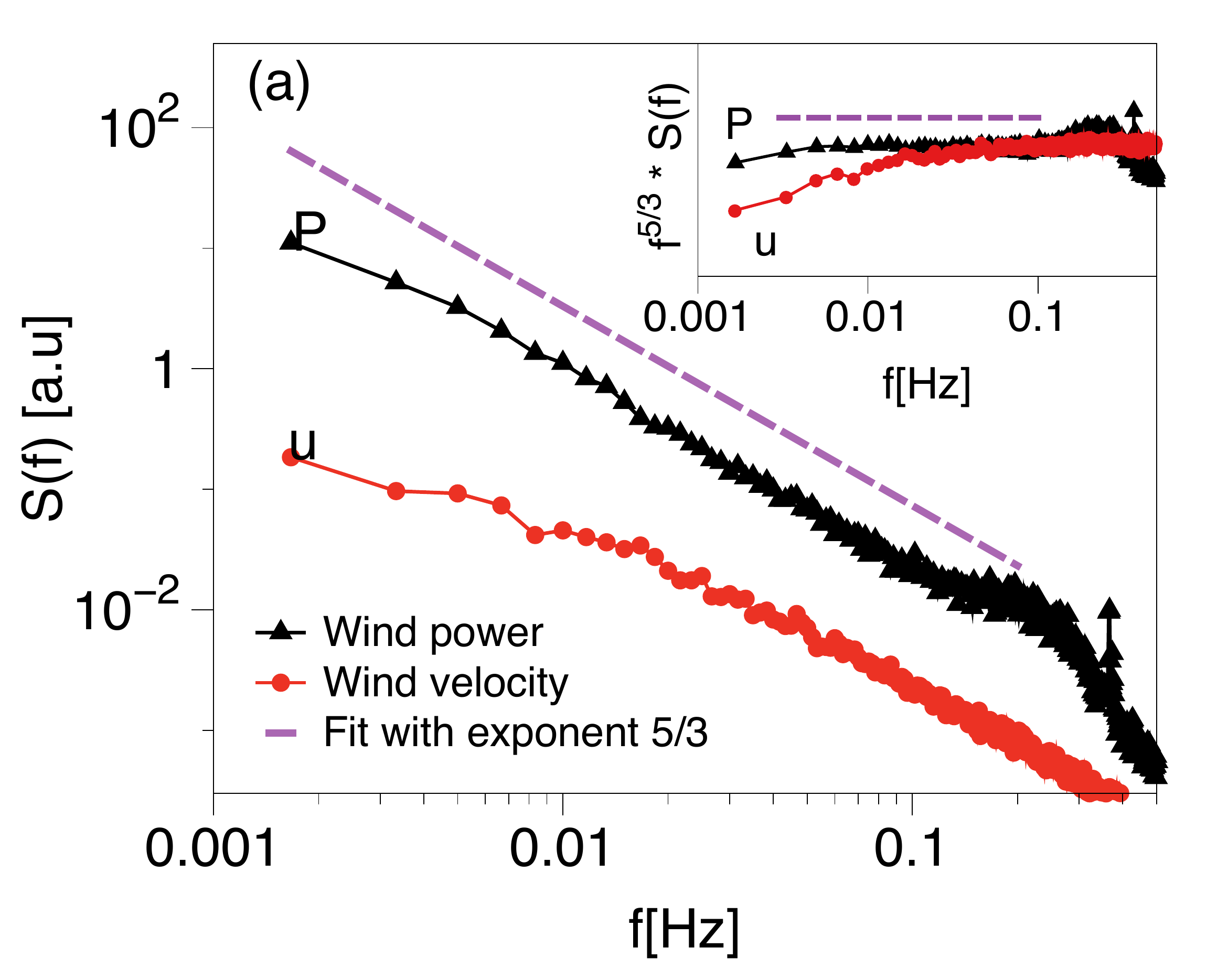}
\includegraphics[width=0.3\hsize]{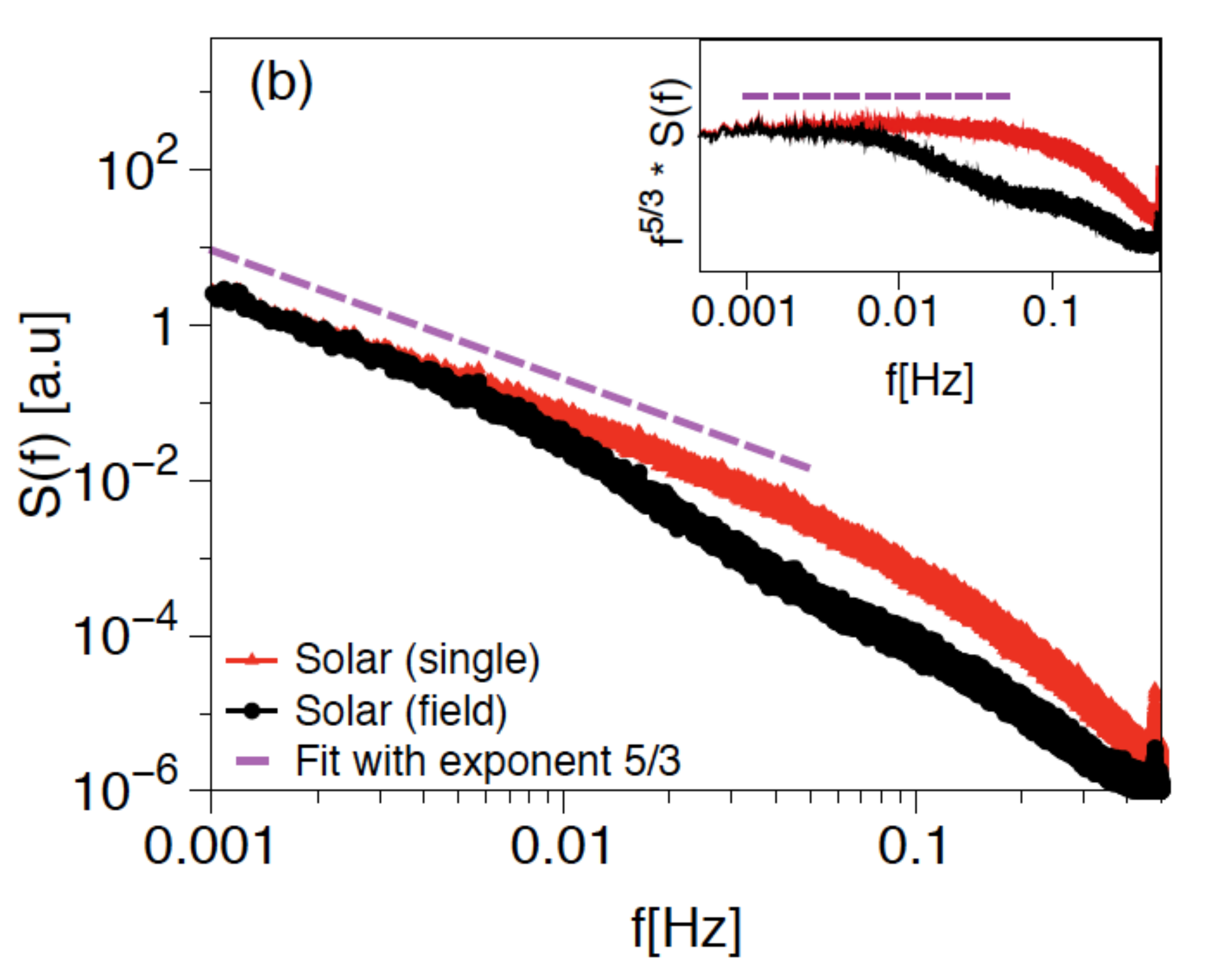}
\includegraphics[width=0.3\hsize]{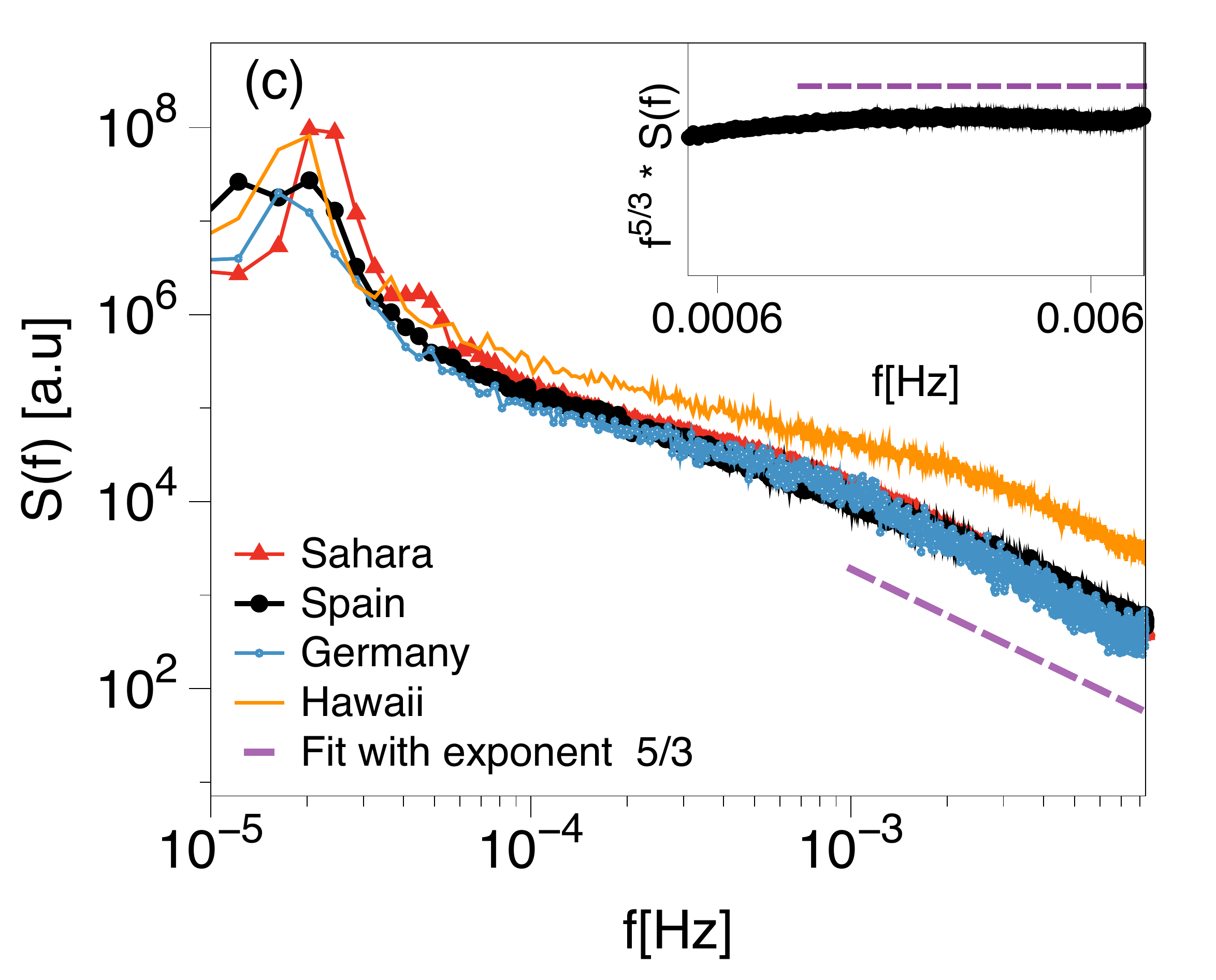}
\caption{ (a)  Power spectra of wind velocity, wind power fluctuations in log-log scale, for a data set with a resolution of $1$~Hz (W1). The Kolmogorov exponent $5/3$ is represented by dashed lines \cite{2,7-3}. (b) Power spectra of irradiance fluctuation for a single site (red) and averaged over 16 sensors (black) in log-log scale measured in Hawaii (S2) with a sample rate of $1$Hz. (c) Power spectra of irradiance fluctuations for minute-averaged solar irradiance in several regions around the world (Hawaii, Sahara, Spain, Germany), again show a turbulence type spectrum $5/3$-law. In the inset of (a), (b) and (c), log-log plots of the compensated energy spectra $f^{5/3} S(f)$ versus frequency $f$ are shown. In the inset of (c) the compensated energy spectrum is plotted for the irradiance in Spain.}\label{3}
\end{figure}

Let us begin with known results about the power spectrum of solar and wind power. The power spectra computed from high frequency time series (with sample rate 1~Hz) of solar irradiance, wind velocity and wind power exhibit a power-law behaviour with an exponent $\sim 5/3$ (Kolmogorov exponent \cite{2,7-2}) in the frequency domain $0.001 < f < 0.1$~Hz, indicating that they are turbulent-like sources \cite{7-2,7-3,7-5}. This is reconfirmed here in Figs. \ref{3}a and \ref{3}b for Germany (W1) and Hawaii (S2), respectively. 
As shown in Fig. \ref{3}b, the fast fluctuations of single sensor measurements are partly filtered in high frequencies for the cumulative irradiance fluctuations of a geographically averaged solar field. A similar filtering effect has been observed also in the cumulative power of wind farms \cite{7-3}. 

Also the power spectra of one minute averaged solar irradiance fluctuations in several regions around the world (S1-S4) for frequencies $ 0.001< f < 1/120$~Hz again show a turbulence-type spectrum $\sim 5/3$-law, as shown in Fig. \ref{3}c, indicating a universal characteristic of the power spectrum. 
The scaling with the same exponent for all measured high frequency time series (to the best of our knowledge first investigated in \cite{7-4}) means that the power grid is being fed by turbulent-like sources.

\begin{figure}
\centering
\includegraphics[width=0.3\hsize]{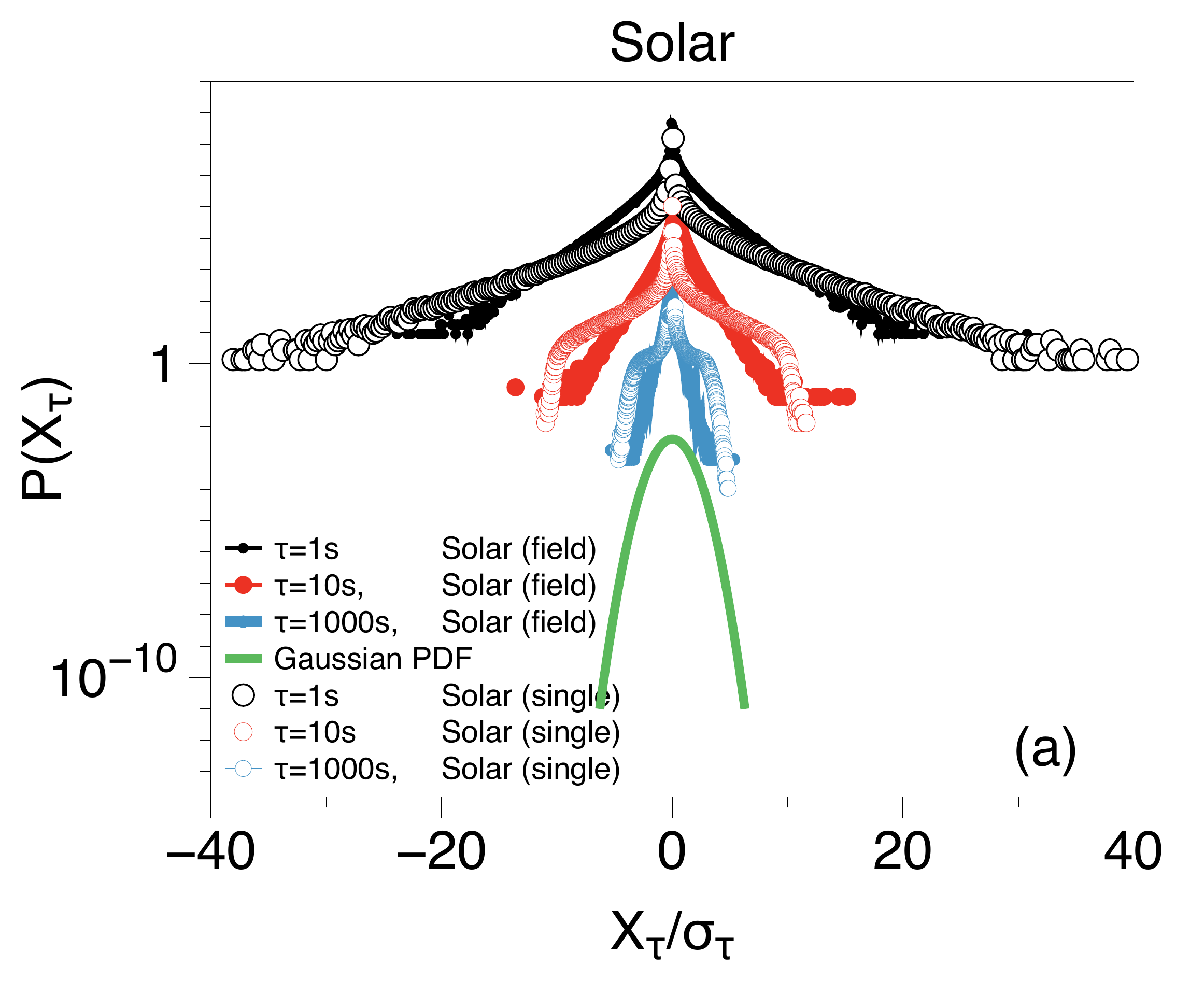}
\includegraphics[width=0.3\hsize]{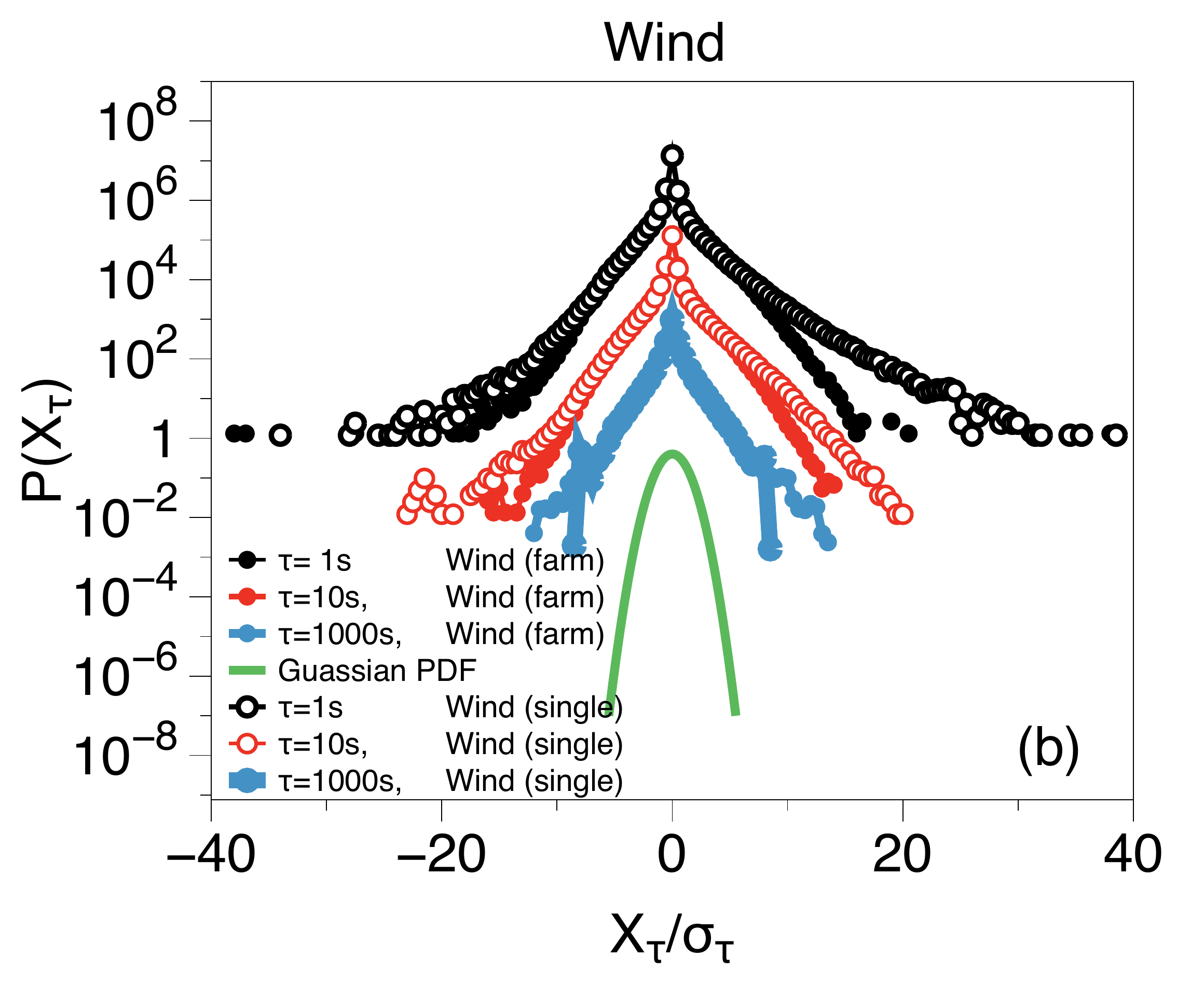}
\includegraphics[width=0.3\hsize]{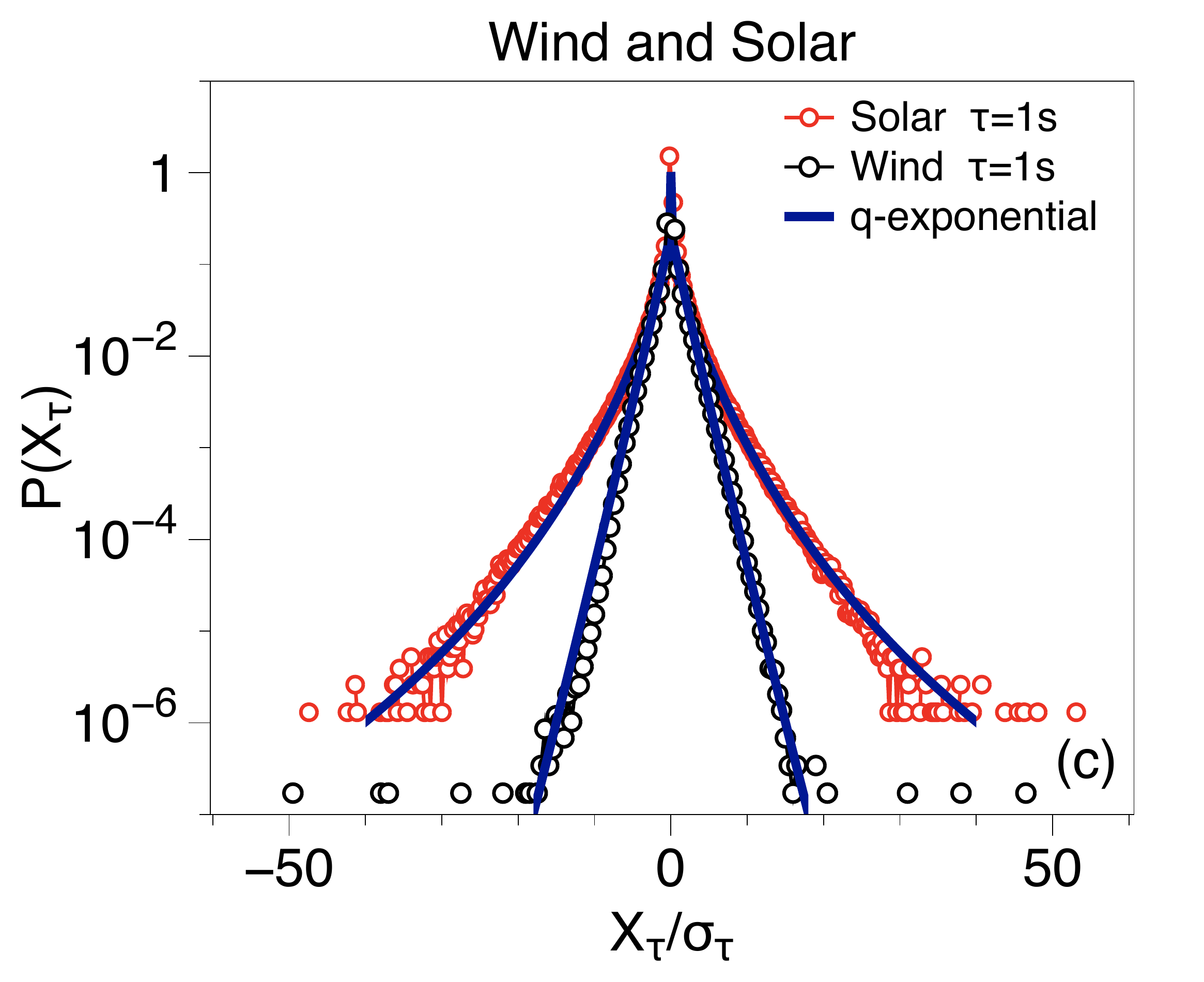}
\caption{Probability distribution functions (PDF) of increment statistics, $P(X_{\tau})$ for solar and wind power fluctuations. a) Continuous deformation of the increment PDFs for time lags $\tau=1,10,1000$ sec in log-linear scale, for the solar irradiance fluctuations of a single sensor and the whole field (S2). The PDFs are shifted in the vertical direction for convenience of presentation and $X_{\tau}$s are measured in units of their standard deviation $\sigma_\tau$. b) Same figure for the increment PDFs of one wind turbine and a wind farm power for the same time lags. A Gaussian PDF with unit variance is plotted for comparison. c) Comparison of the increment PDFs of wind and solar power time series having a similar rated power with time lag $1$~s. Solid curves are fits based on q-exponential functions Eq. (\ref{qexp}). The obtained parameters are $\beta=0.64$, $q=1.12$ for solar and $ \beta=0.87$, $q=1.01$ for wind power PDFs. The dot size is chosen in the order of the statistical error.}\label{4}
\end{figure}

Next we study the shapes of increment PDFs $ P(X_{\tau})$, normalised to their standard deviations, expanding the above analysis of the $\tau$-dependence of increment PDFs standard deviation by the power spectrum. Results of solar irradiance data (S2) and wind power time series (W1) are shown in Fig. \ref{4}a and Fig. \ref{4}b for the time lags $\tau=1,10,1000$~s. The normalised increment PDFs depart largely from the normal (Gaussian) distribution, as they possess exponential-like fat tails. These tails extend to extreme values like 20 $\sigma_{\tau=1s}$ and more. As such events would not be expected from normal probability we refer to them as "extreme events". From Fig. \ref{4}a and Fig. \ref{4}b, it becomes clear how these increment statistics change with the scale $\tau$.

Figs. \ref{4}a and \ref{4}b depict that not only the increment PDFs of the single wind turbine and the single solar sensor depart largely from the normal distribution, but also the wind farm and solar field deviate significantly from the Gaussian distribution. For instance, 20 $\sigma_{\tau=1s}$ fluctuations are observed on average once a month for wind power data (W1), and $\sim 1000$ times per month for solar irradiance (S2). Characterising these data, as often done, only by the variance or power spectra, and assuming a Gaussian process, such extreme events would be expected only once every 3 million years. Hence, it is worth to emphasise that, if instead of intermittent PDFs, common Gaussian-distributed processes are used for grid stability studies, these extreme events will not be taken into account, which can cause unrealistic results for grid stability analyses.

To compare the characteristics of solar and wind power production, we present in Fig. \ref{4}c the power increment statistics for two units with the same rated power. 
Wind power features extreme events up to about 20 $\sigma_{\tau=1}$, while up to about 40 $\sigma_{\tau=1}$ are recorded for solar irradiance in this time lag. 
The probability of observing 20 $\sigma_{\tau}$ fluctuations of solar irradiance in $1$~s is three orders of magnitude higher than that of wind power. 
Solar irradiance thus has much more frequent extreme events, which is again an important aspect for gird integration. 

Note that in this study solar power systems are different from wind power, as they are represented by analyses of solar irradiance and not by solar electric power. This is justified by the direct and quasi-linear transformation of plane-of-array solar irradiance into solar power, assuming horizontally oriented modules in this case.  Any deviations from this behaviour due to the physical characteristics of both solar cells and additional system components (e.g.\ inverter) are small and thus neglected in this study. Especially, due to the extreme fast response of PV systems to irradiance, they perfectly reproduce any intermittent pattern in the irradiance time series. Statistical characteristics derived from solar irradiance time series are therefore valid also for solar power time series with high accuracy. 

\begin{figure}
\centering
\includegraphics[width=0.4\hsize]{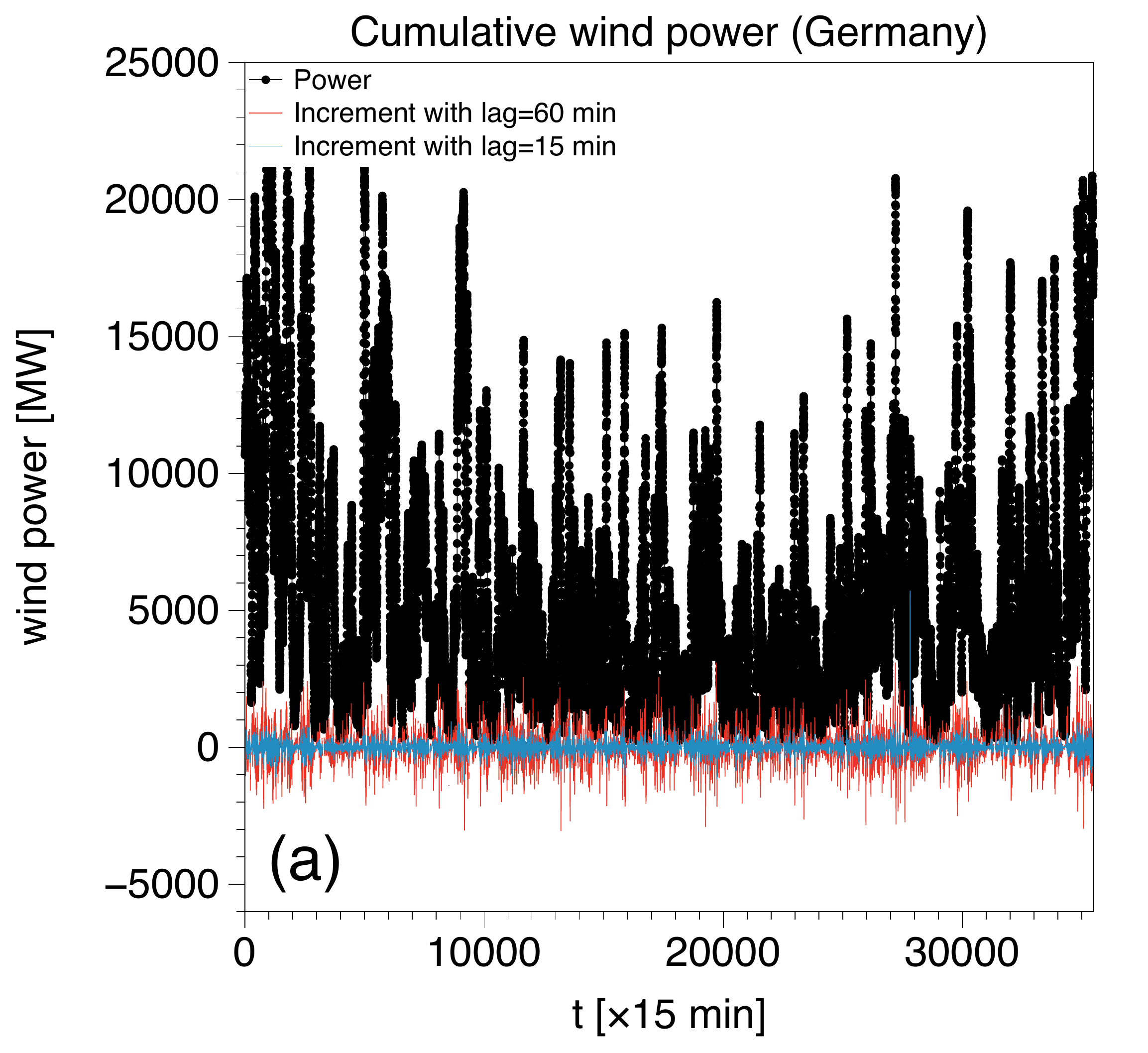}
\includegraphics[width=0.4\hsize]{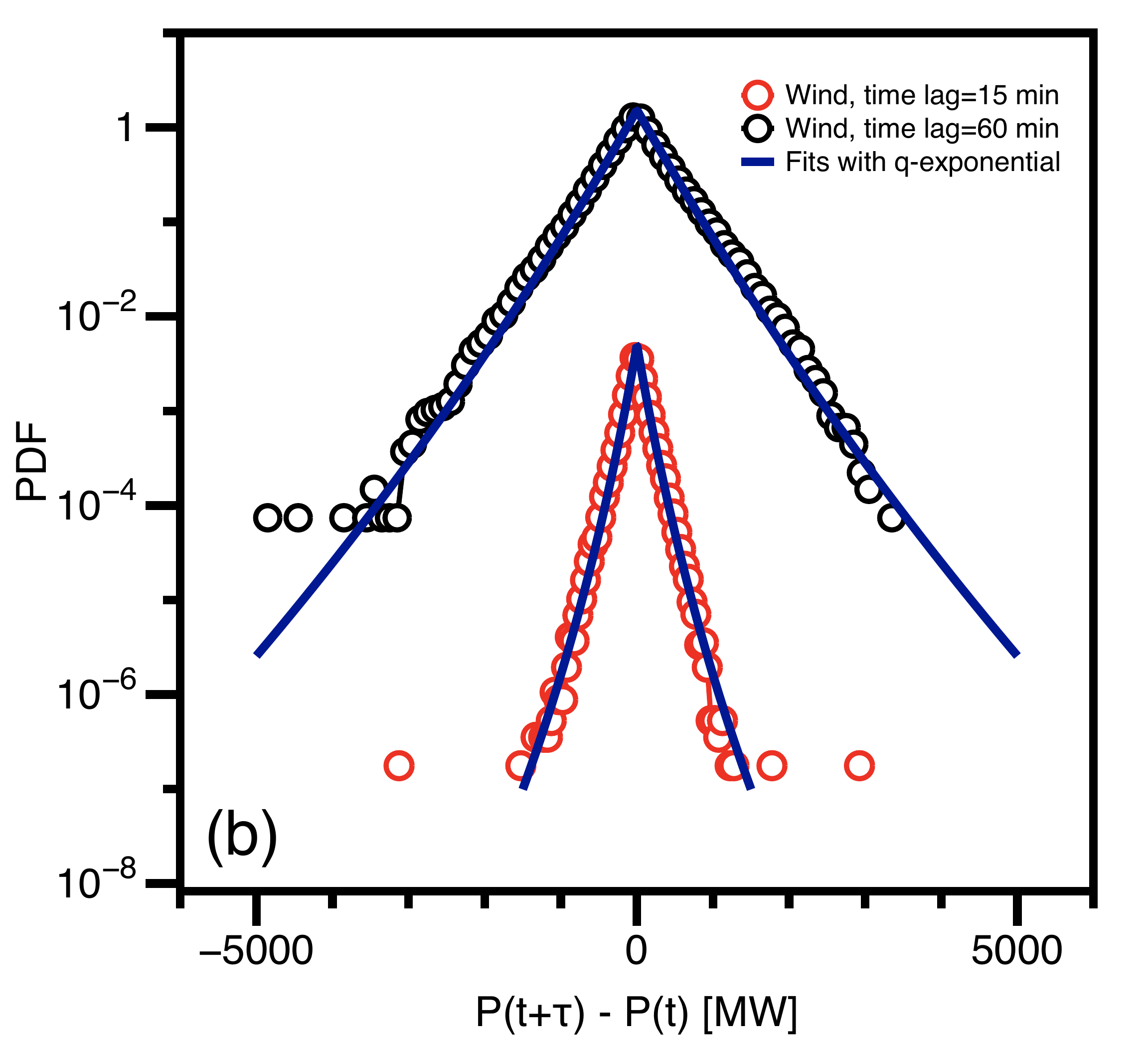}
\caption{(a) Total wind power output and its increments in time lags $15$~min and $1$~h in Germany for the year 2012, showing a strongly intermittent behaviour. The installed capacity is about $\sim 30$~GW. (b) Deformation of the increment PDFs for time lags $\tau=15, 60$~min in log-linear scale, for wind power in Germany (with a rated power $ \sim 30 $~GW). Extreme events up to about $\pm 2000$~MW and  $\pm 4000$~MW are recorded in time lags 15 min and 60 min, in Germany respectively. Solid curves are fits based on q-exponential functions Eq. (\ref{qexp}).  For wind power in Germany the obtained parameters are $\beta=0.003$, $q=1.03$ and $\beta=0.009$, $q=1.02$ for time lags $\tau=1$h and $\tau=15$min, respectively.}\label{6}
\end{figure}

\begin{figure}
\centering
\includegraphics[width=0.4\hsize]{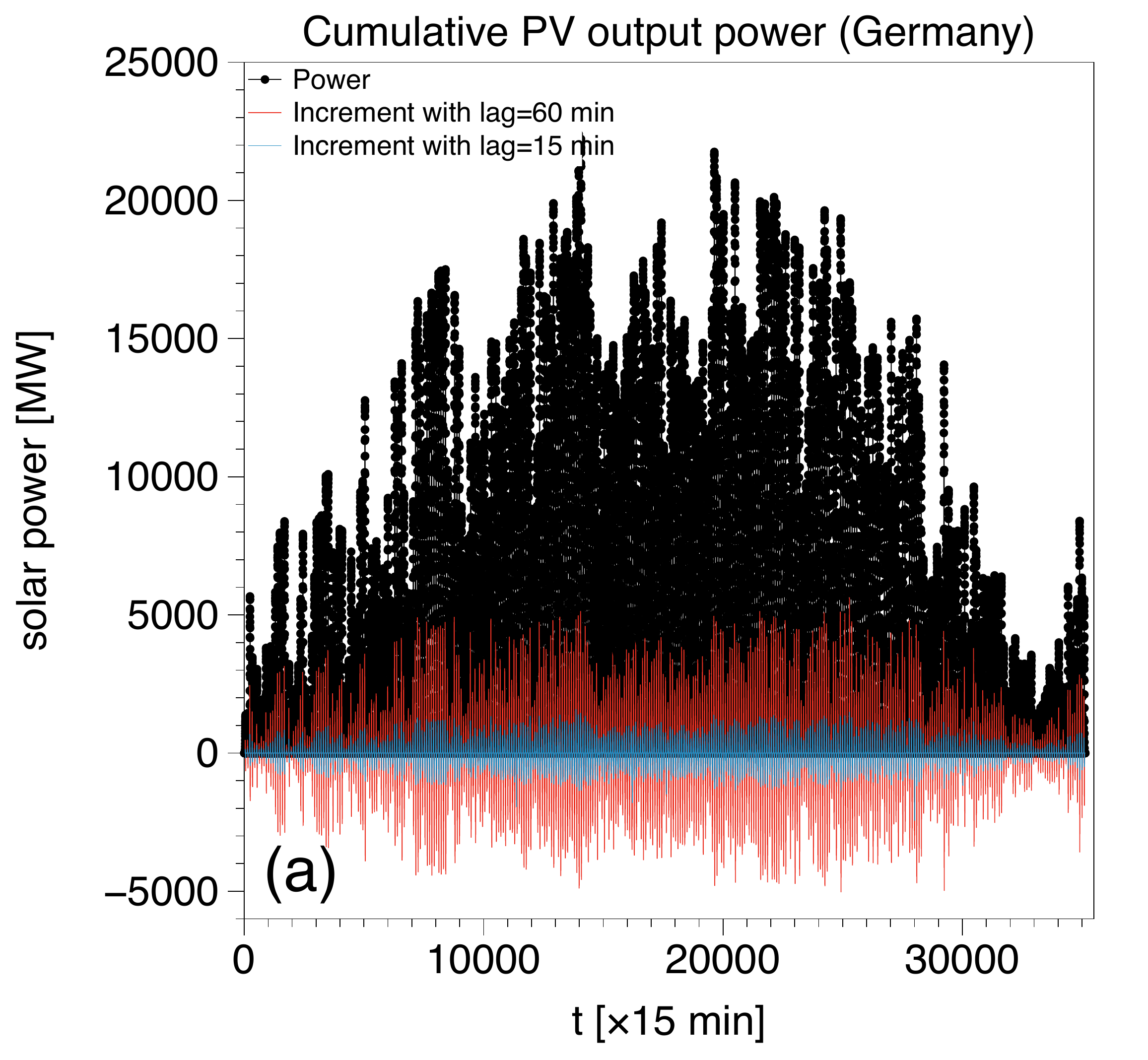}
\includegraphics[width=0.4\hsize]{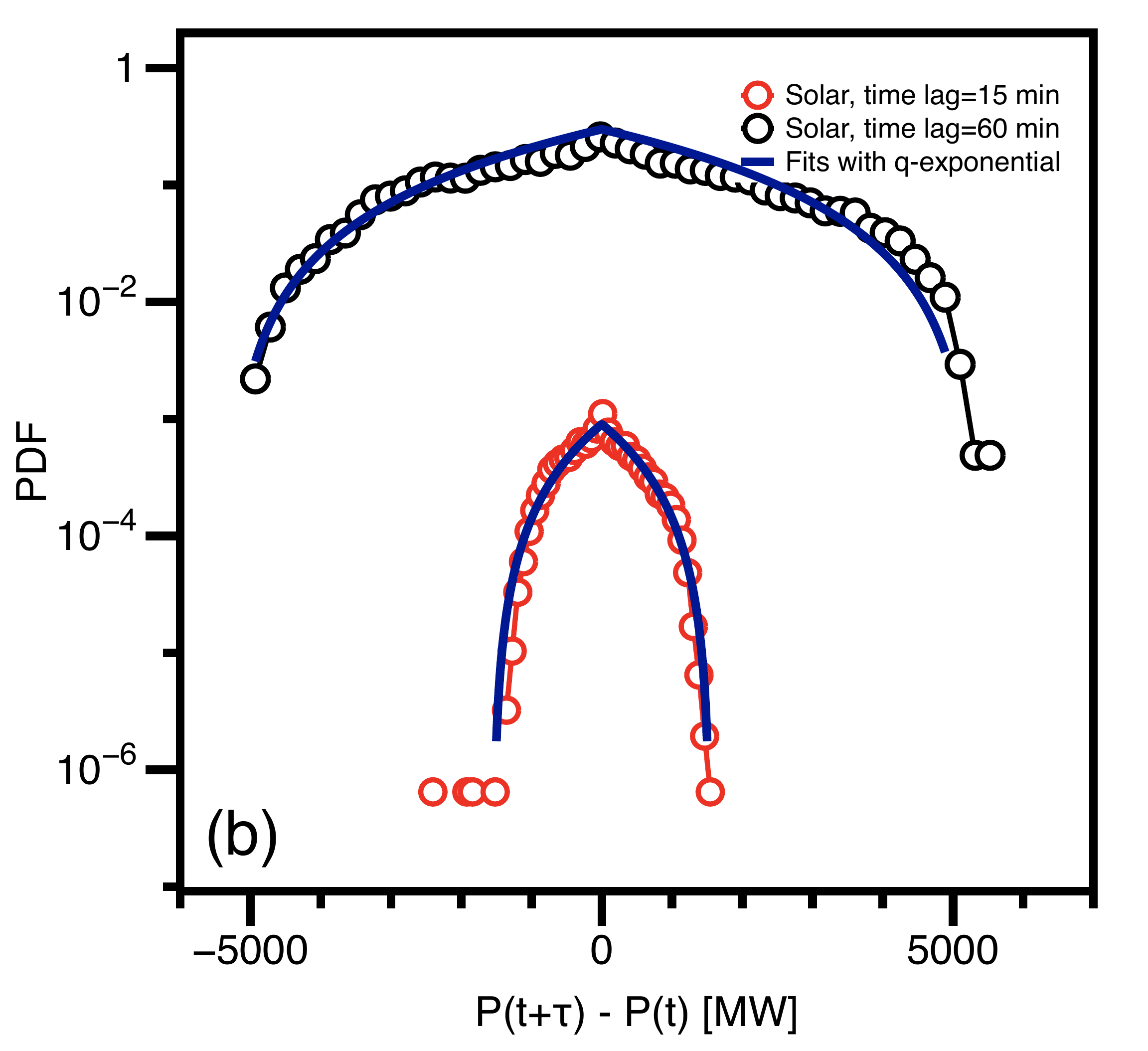}
\caption{(a) Total solar power output and its increments in time lags $15$~min and $1$~h in Germany for the year 2012, showing strong variability. The installed capacity is about $\sim 30$~GW. (b) Deformation of the increment PDFs for time lags $\tau=15, 60$~min in log-linear scale, for solar power in Germany (with a rated power $ \sim 30 $~GW). For a 60 min time lag, extreme events up to  $\pm 6000$~MW are recorded in cumulative PV output in Germany. Solid curves are fits based on q-exponential functions Eq. (\ref{qexp}).}
\label{1}
\end{figure}

Now let us study the non-Gaussian properties of the increment statistics of renewable wind and solar power from nationwide installations. Typical time series of aggregated wind and solar power in Germany (and their increments) are given in Figs.~\ref{6}a and \ref{1}a, showing very strong variability and fluctuations.

\begin{figure}
\centering
\includegraphics[width=0.4\hsize]{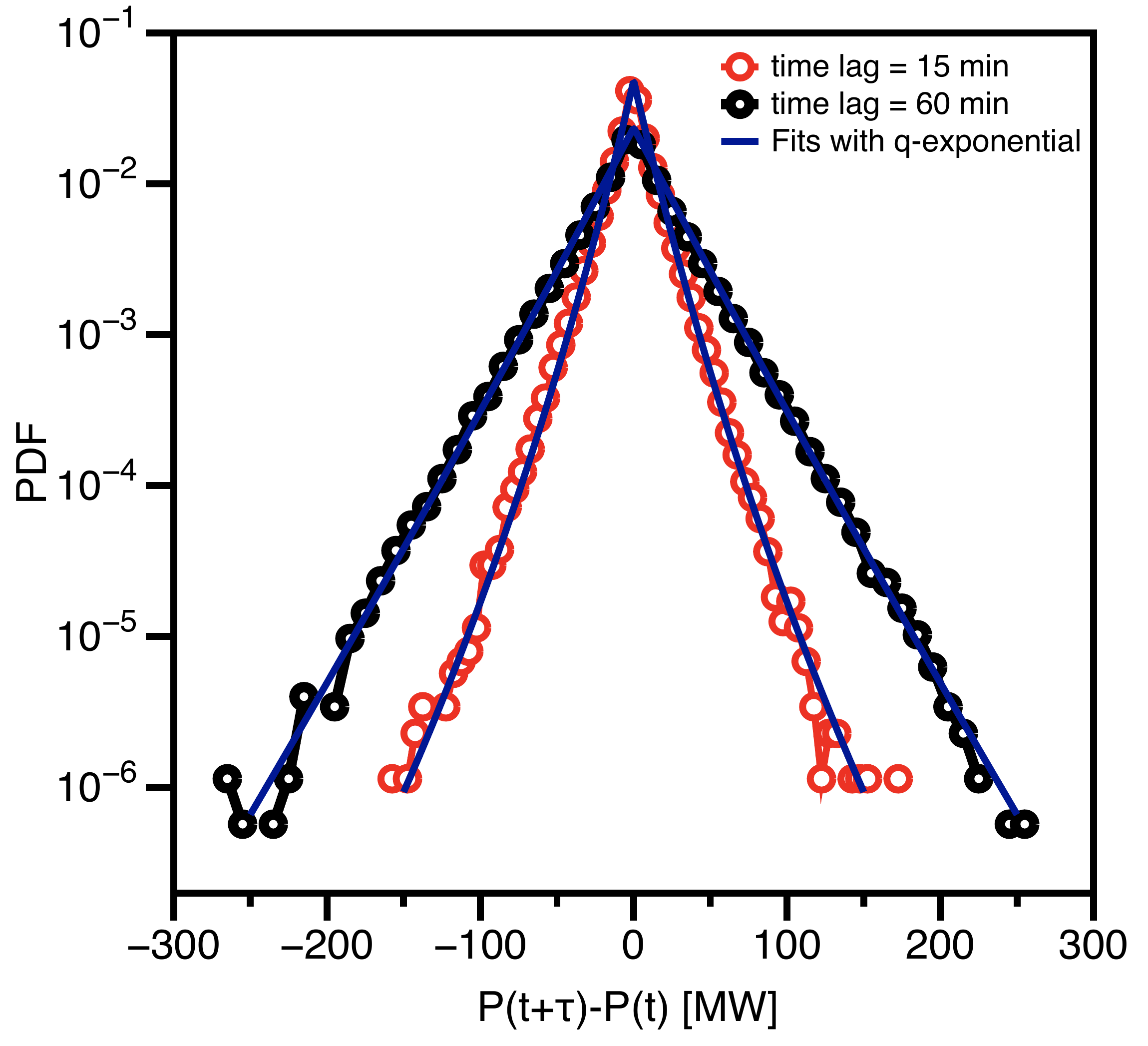}
\caption{Deformation of the increments PDFs for time lags $\tau=15, 60$ min in log-linear scale, for wind power in Ireland (with a rated power $ \sim 1$$ GW$). Extreme events up to about $\pm 150$$ MW$ and $\pm 300 $$MW$ are recorded in time lags 15 min and 60 min, in Ireland respectively. Solid curves are fits based on q-exponential functions Eq. (\ref{qexp}). For wind power in Ireland the obtained values are  $\beta=0.102$, $q=1.06$ and $\beta=0.0466$, $q=1.02$ for time lags $\tau=1$h and $\tau=15$min, respectively.}
\label{2}
\end{figure}

In Figs.~\ref{6}b and~\ref{1}b and Fig.~\ref{2}, increment PDFs for time lags $\tau=15, 60$~min are shown for aggregated wind and solar power in Germany (both sources with a rated power $\sim 30$~GW), and for wind power in Ireland (with a rated power $ \sim 1$~GW), see also \cite{15}. As a remarkable result, the non-Gaussian characteristics remain for the aggregated power output of country-wide installations.
Ramp events up to about $\pm 2000$~MW ($\pm 150$~MW) and  $\pm 4000$~MW ($\pm 300 $~MW) are recorded for 15 and 60 minute time lags in Germany (Ireland). This is a direct consequence of the long-range correlations of wind velocity and cloud size distributions that are $\sim 600$ km and $\sim 2100$ km, respectively \cite{16,17}. Therefore, the central-limit theorem, predicting a convergence to Gaussianity, does not apply. Note also that in Fig.~\ref{1}b the probability of observing $\pm 4000$~MW fluctuations of solar power in $60$~min is two orders of magnitude higher than that of wind power for nearly the same rated power in Germany.

\begin{figure}
\centering
\includegraphics[width=0.5\hsize]{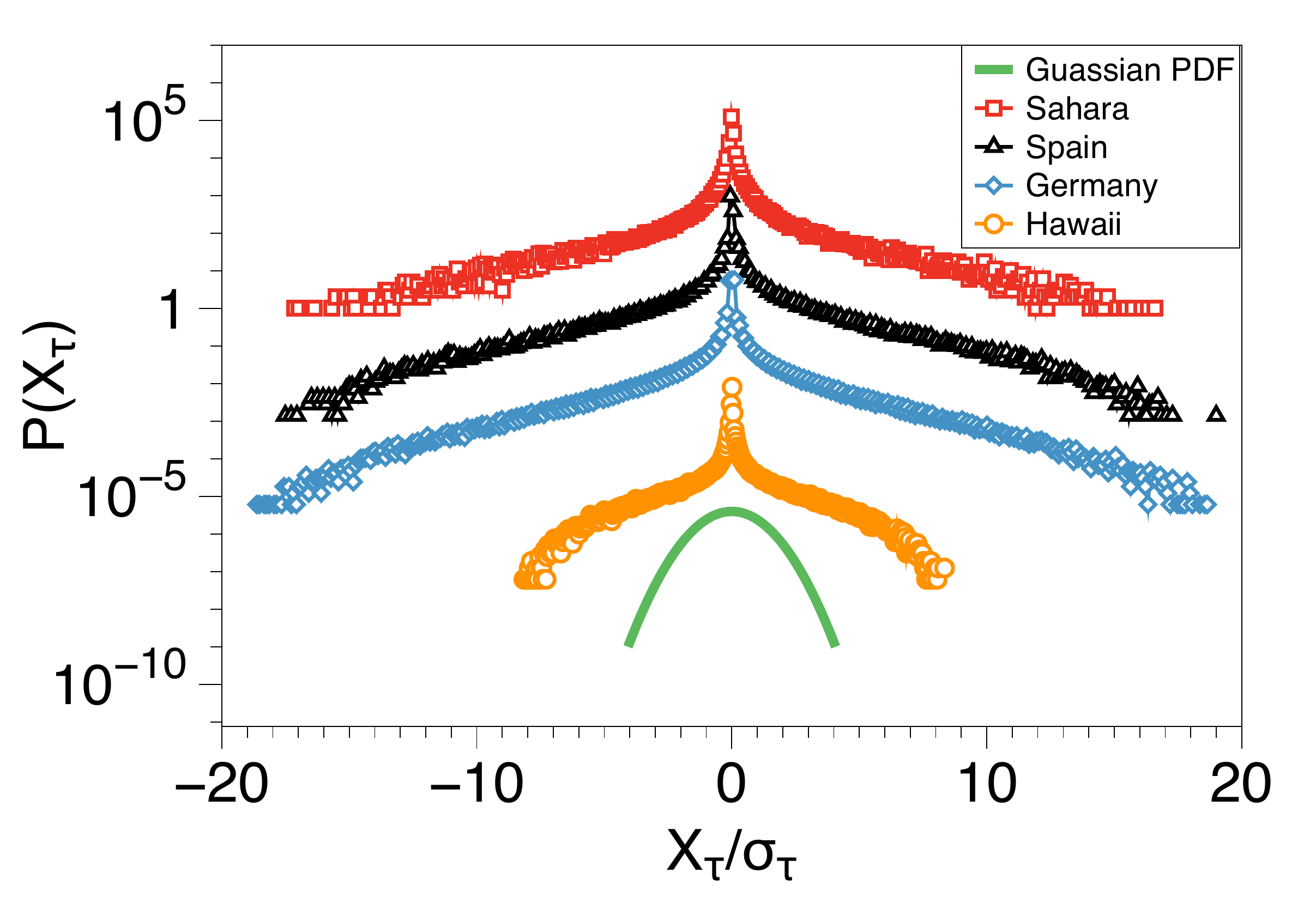}
\caption{ Probability distribution functions (PDF) of increment statistics $P(X_{\tau})$ in log-linear scale for a time lag of 1~min, based on minute-averages of solar irradiance in several regions around the world (S1-S4). The PDFs are shifted in the vertical direction for convenience of presentation and $X_{\tau}$s are measured in units of their standard deviation $\sigma_\tau$. }\label{5}
\end{figure}

For further investigation, Fig.\ref{5} depicts the increment PDFs of solar irradiance in several regions around the world, based on one minute averaged data (S1-S4) and a corresponding time lag of 1~min. These data sets exhibit similar non-Gaussian characteristics, with extreme events up to about 10-20 $\sigma_{\tau=1\,min}$ having been recorded.

\begin{figure}
\centering
\includegraphics[width=0.5\hsize]{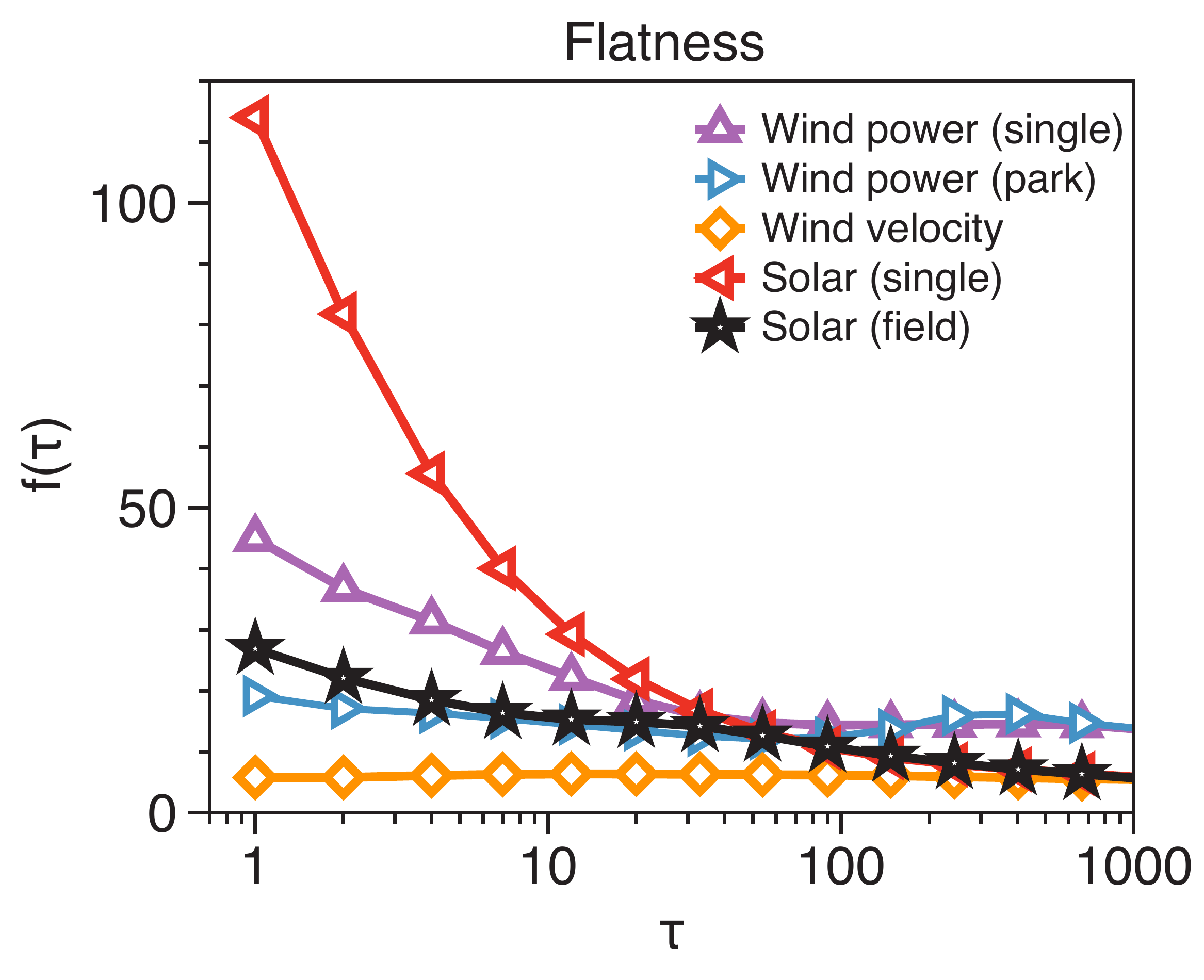}
\caption{The lag-dependence of the flatness $ f(\tau) = S_4(\tau) / S_2 (\tau) ^2$, (where $S_{2k} = \langle ( X (t+ \tau)-X (t) )^{2k} \rangle $) for solar irradiance, wind power and wind velocity fluctuations. They deviate strongly from the value $3$ that corresponds to a Gaussian distribution, especially on short time scales.}\label{7}
\end{figure}

To quantify the time scale dependence of the intermittency, the lag-dependence of the flatness, 
is shown in Fig.~\ref{7} for the wind velocity as well as for the wind power and solar irradiance. The flatness increasingly deviates from the value $3$ (which corresponds to a Gaussian distribution) on short time scales. For the time lag $\tau=1$~s, the flatness reaches values $30-120$ for solar irradiance data, $20-40$ for wind power data and $6$ for wind velocity. The results for the flatness quantitatively confirm the findings from the PDF study as discussed above. Intermittency decreases on larger time scales and with averaging over more units, but stays above the Gaussian limit. Fig. \ref{7} shows that the flatness, and hence non-Gaussianity, is larger for solar irradiance than for wind power on time scales $< 1$~min and becomes smaller for $> 1$~min. We would like to stress that the increments are strongly correlated on short time scales, see \cite{njpcor} for a recent discussion.

For the practical purpose of predicting the likelihood of large power fluctuations, we parametrise the intermittent shape of the increment PDFs using the $q$-exponential function \cite{Tsallis1988}

\begin{equation}
P(X_{\tau})= A~[1-\beta~(1-q)~|X_{\tau}|~]^{1/(1-q)},
\label{qexp}
\end{equation}
with fitting parameters $\beta$ and $q$, and normalisation constant $A=1/2 (2-q) \beta$.
As shown in Figs. \ref{4} and \ref{6} this model fits the observed PDFs of normalised increments $X_{\tau} / \sigma_{\tau}$ very well. 
It is straightforward to show that the relation between flatness $f$ and parameter $q$ in lag $\tau$ is:

\begin{equation}
f(\tau)= 6 ~\frac{(2q(\tau)-3)(3q(\tau)-4)}{(4q(\tau)-5)(5q(\tau)-6)} 
\end{equation}
and that $q$ can be expressed in terms of the flatness (for $f \geq 2.4$) as
\begin{equation}
q(\tau)= - ~ \frac{\sqrt{f(\tau)^2+84f(\tau)+36}-49f(\tau)+102}{40f(\tau)-72}.
\label{qq}
\end{equation}
As we see from Eq. (2), the flatness is independent of parameter $\beta$. For a given lag, we can first calculate the parameter $q$ from its flatness and then parameter $\beta$ can be evaluated via variance, i.e. $<X_{\tau}^2>= \{2(q-2)\} / \{ \beta^2(-4+3q)(6-7q+2q^2) \}$ (to find the parameters $q$ and $\beta$ we can also use a minimisation of distance between experimental increment PDFs and q-exponential, as in Figs. \ref{4} and \ref{6}). 

\begin{figure}
\centering
\includegraphics[width=0.4\hsize]{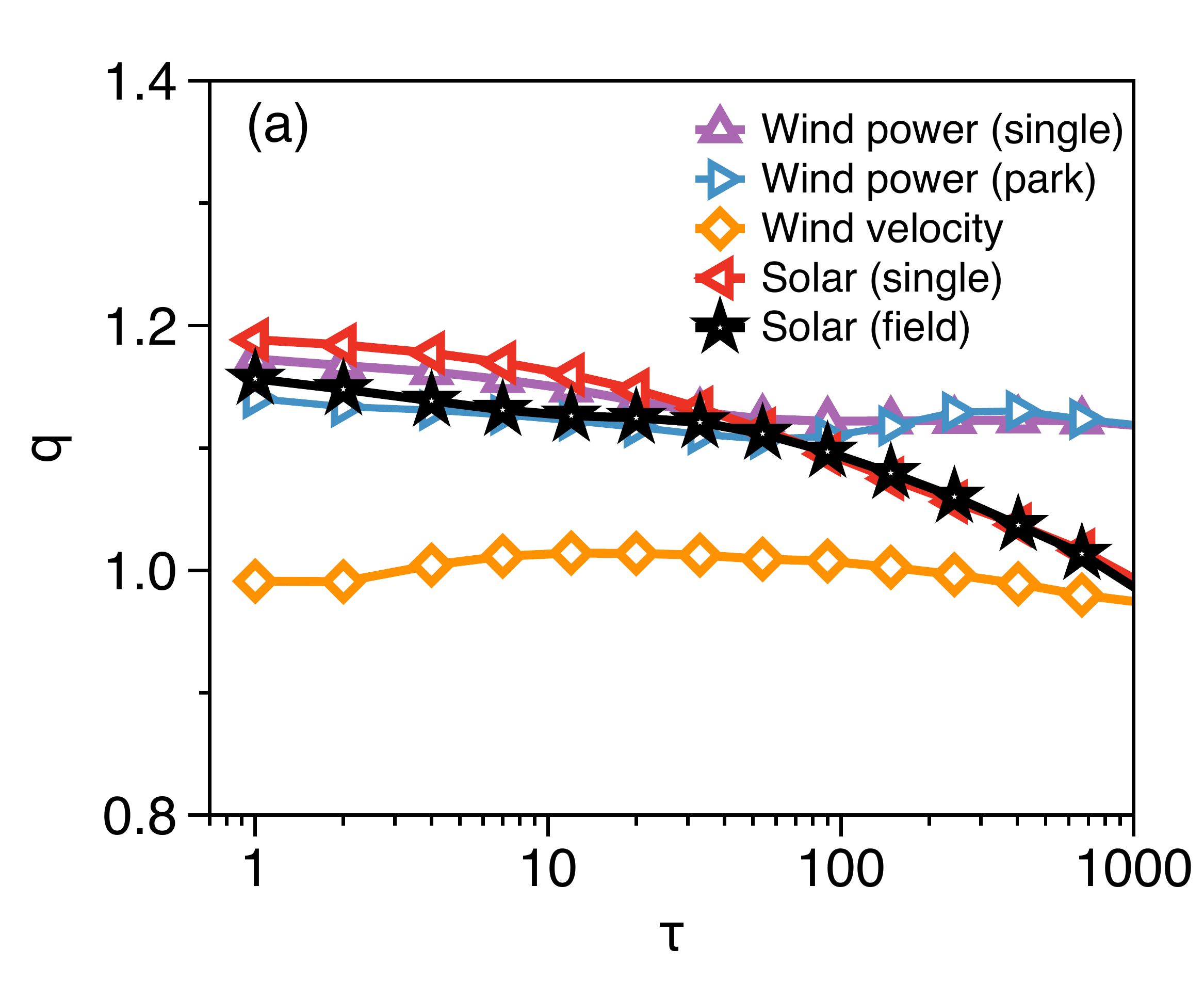}
\includegraphics[width=0.4\hsize]{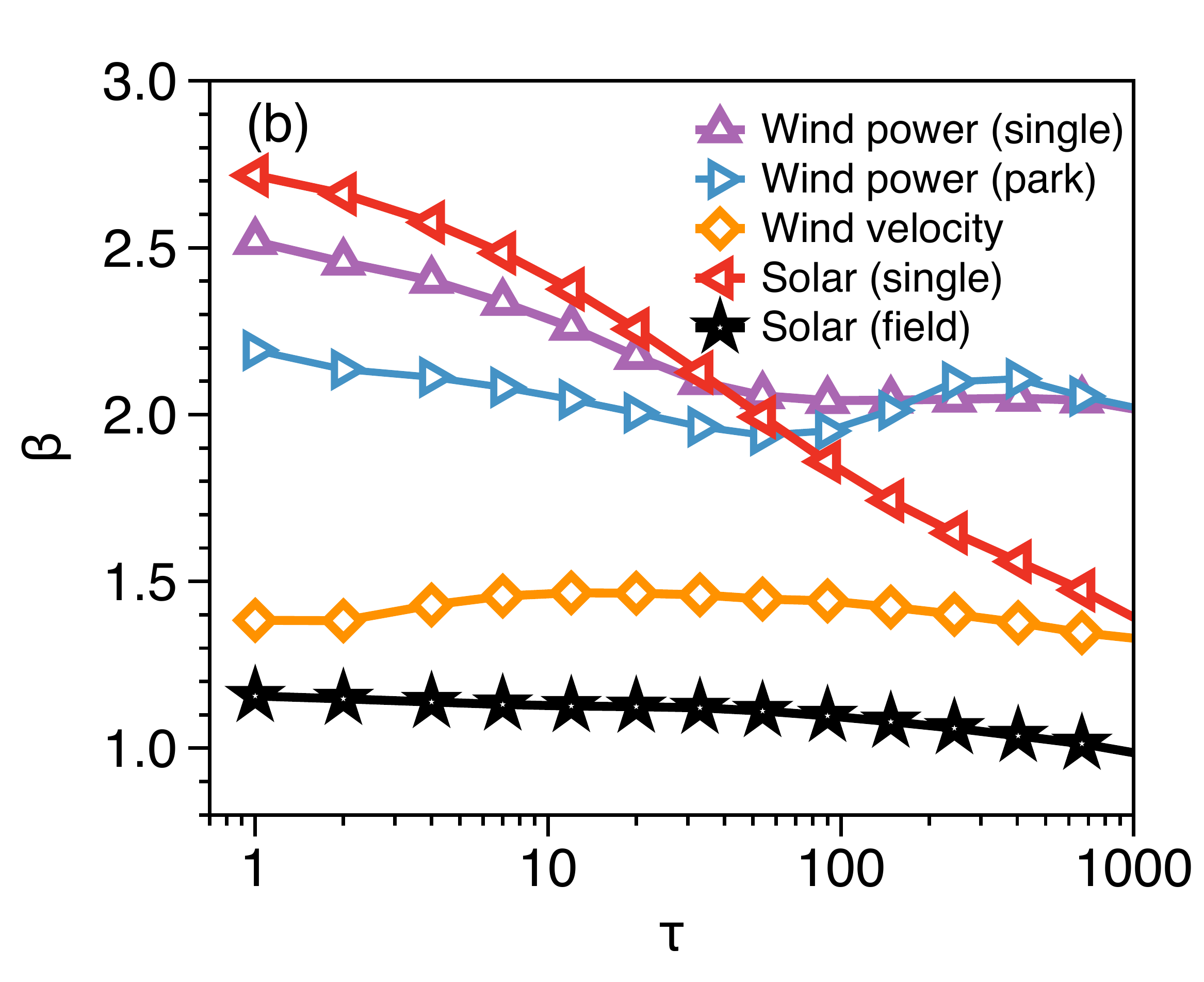}
\caption{The lag-dependence of (a) $q$ and (b) $\beta$, for solar irradiance (S2) and wind power (W1) PDFs in Fig. \cite{4}.}\label{8}
\end{figure}

The $\tau$-dependencies of $\beta$ and $q$ are shown in Fig. \ref{8} for the data sets W1 and S2. For instance, for wind power from a single turbine (data set W1) we find  $ \beta=0.87$, $q=1.01$ for $\tau=1s$, and $\beta=1.15$, $q=1.04$ for $\tau=10$~s in Fig. \ref{4}b. We can conclude that the extreme events statistics of wind and solar power can be very well characterised by $q$-exponential functions for a vast range of $X_{\tau} / \sigma_{\tau}$ values. These results can be used as a basis for stochastic modelling such intermittent time series.

\begin{figure}
\centering
\includegraphics[width=0.5\hsize]{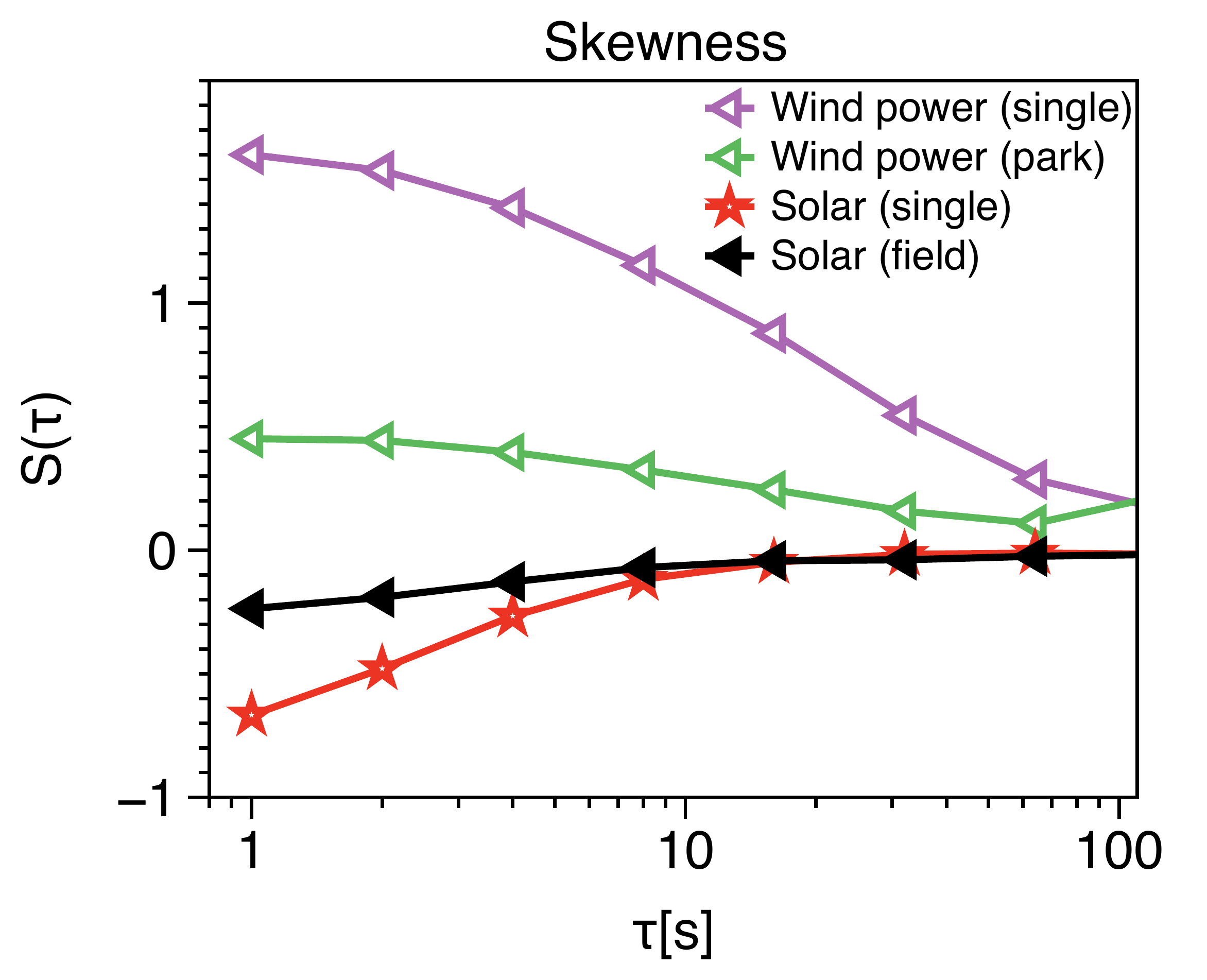}
\caption{The lag-dependence of the skewness  $ S(\tau) = S_3(\tau) / {S_2 (\tau)}^{3/2}$, for solar irradiance and wind power fluctuations. On short time scales, they deviate strongly from zero, which corresponds to a symmetric distribution. Wind power (W1) and solar irradiance (S2) have positive and negative skewness, respectively.}\label{9}
\end{figure}

As specified in Eq. (1), the absolute value of $X_{\tau}$ has been used in the $q$-exponential function, which means that symmetric increment PDFs are assumed for these calculations. 
We should note that the question of symmetric increment distribution is important, as for ideal turbulent
signals a pronounced skewness is expected.
To quantify asymmetric effects in the statistics of positive and negative power increments, the lag-dependence of the skewness is shown in Fig.~\ref{9} for both wind power (W1) and solar irradiance (S2). The lag-dependence of the skewness shows that they deviate in short time scale from zero, which corresponds to a symmetric distribution. 
Wind (solar) power exhibits positive (negative) skewness values, corresponding to a higher (lower) probability of ramp up events than ramp down events.
The skewness of country-wide installations, such as W2, W3, and S5 data sets, is much closer to zero yet. 
Thus we can take the skewness effect as a minor additional contribution to the form of the PDFs, justifying the $q$-exponential form fits as the major one.
This agrees with the good fits to the empirical PDFs shown in Fig.~\ref{4}, \ref{6}, \ref{1} and \ref{2}. 

So far, we have presented a profound characterisation of the power fluctuation statistics measured by increments, with all data showing strong intermittency. 
Details of absolute values of the characterising parameters like the exponent $q$ will change with data sets and seasonal periods of time. 
It will also be worthwhile to see if the estimation of $q$ by the flatness is sufficient to get the best fit, or if it is better to use a free parameter fit for the tails of the PDFs.

\section{Critical transitions at tipping points for dynamics of solar field and wind farm} 

\begin{figure}
\centering
\includegraphics[width=0.4\hsize]{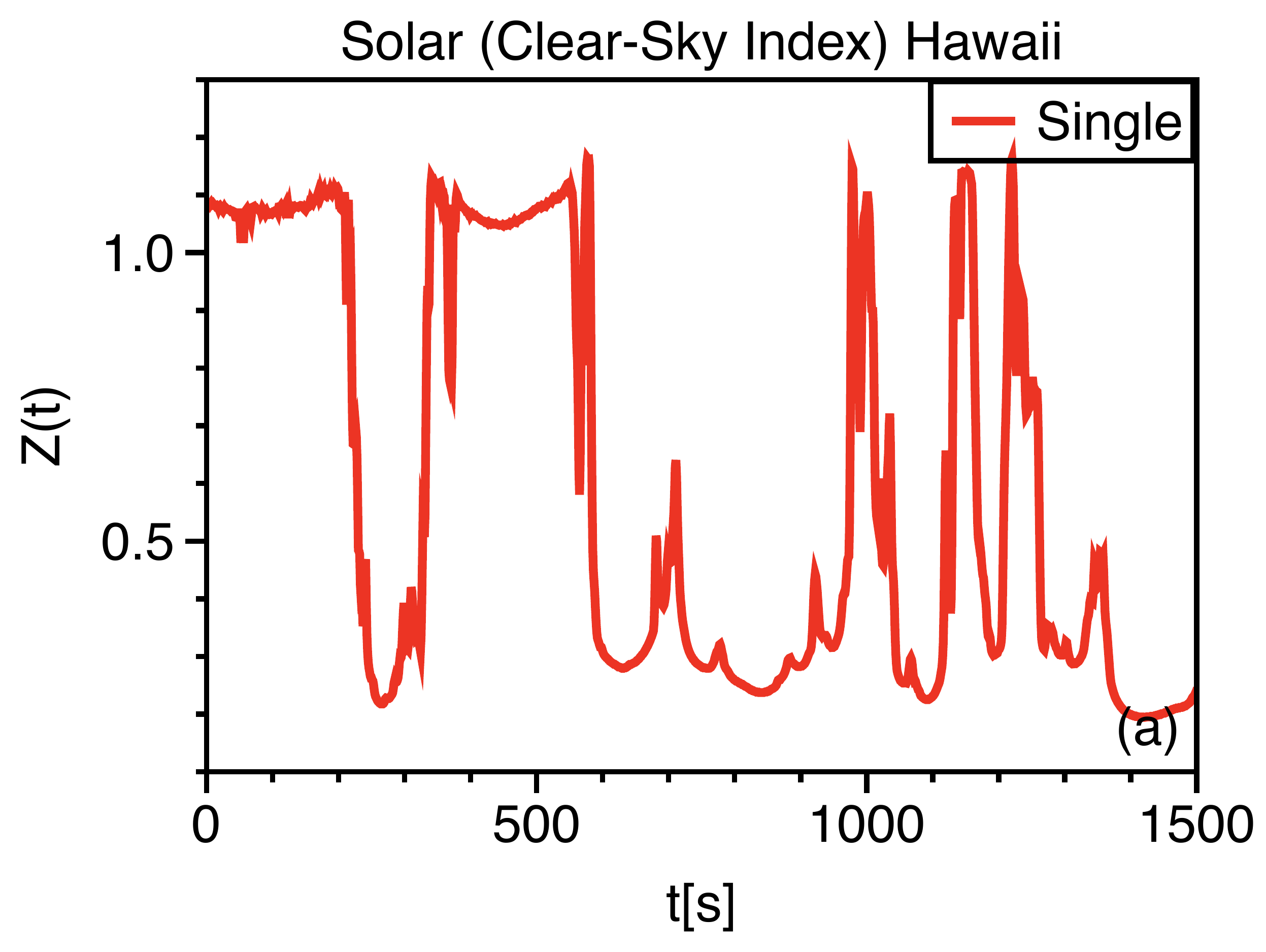}
\includegraphics[width=0.4\hsize]{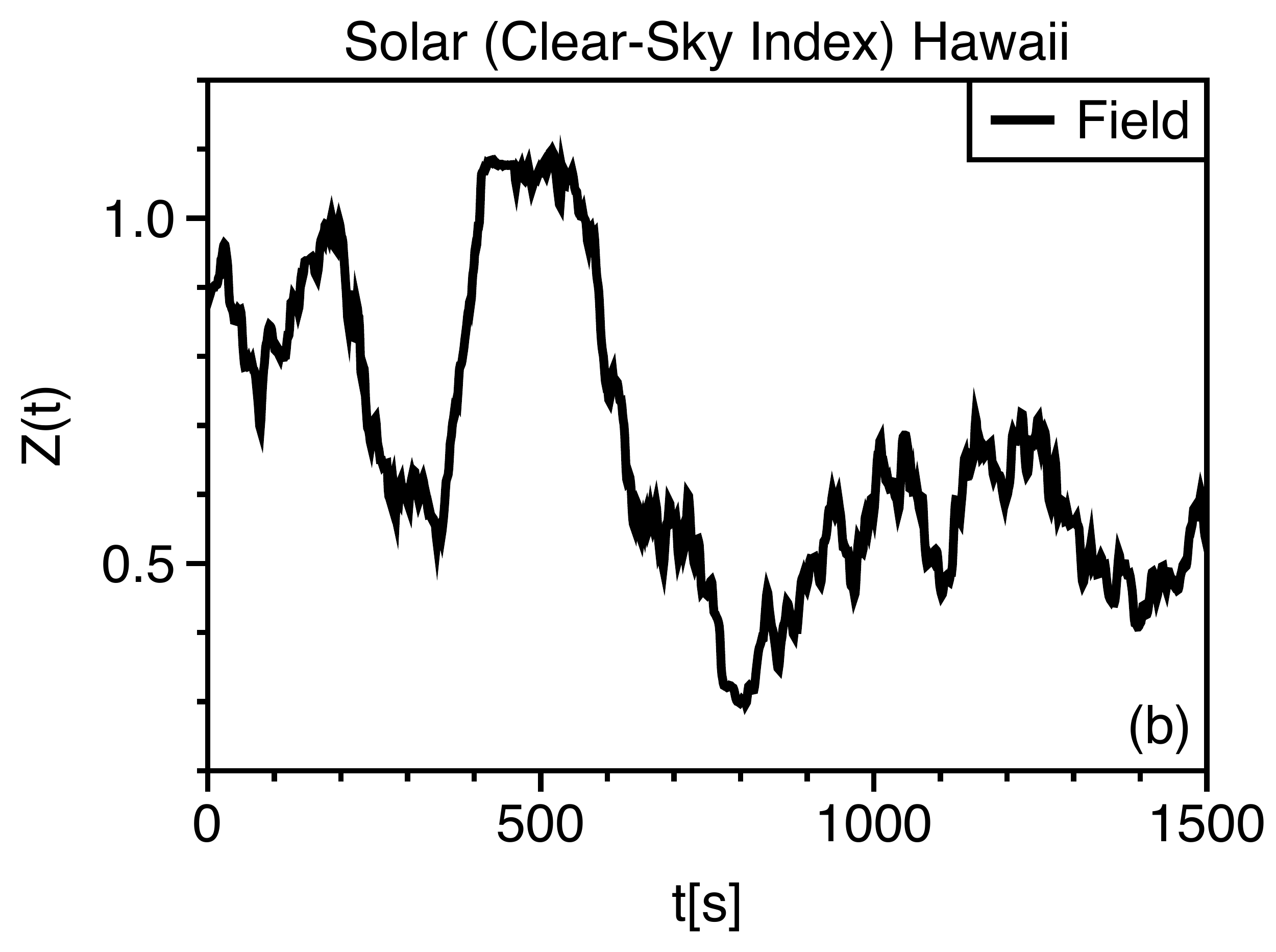}
\includegraphics[width=0.4\hsize]{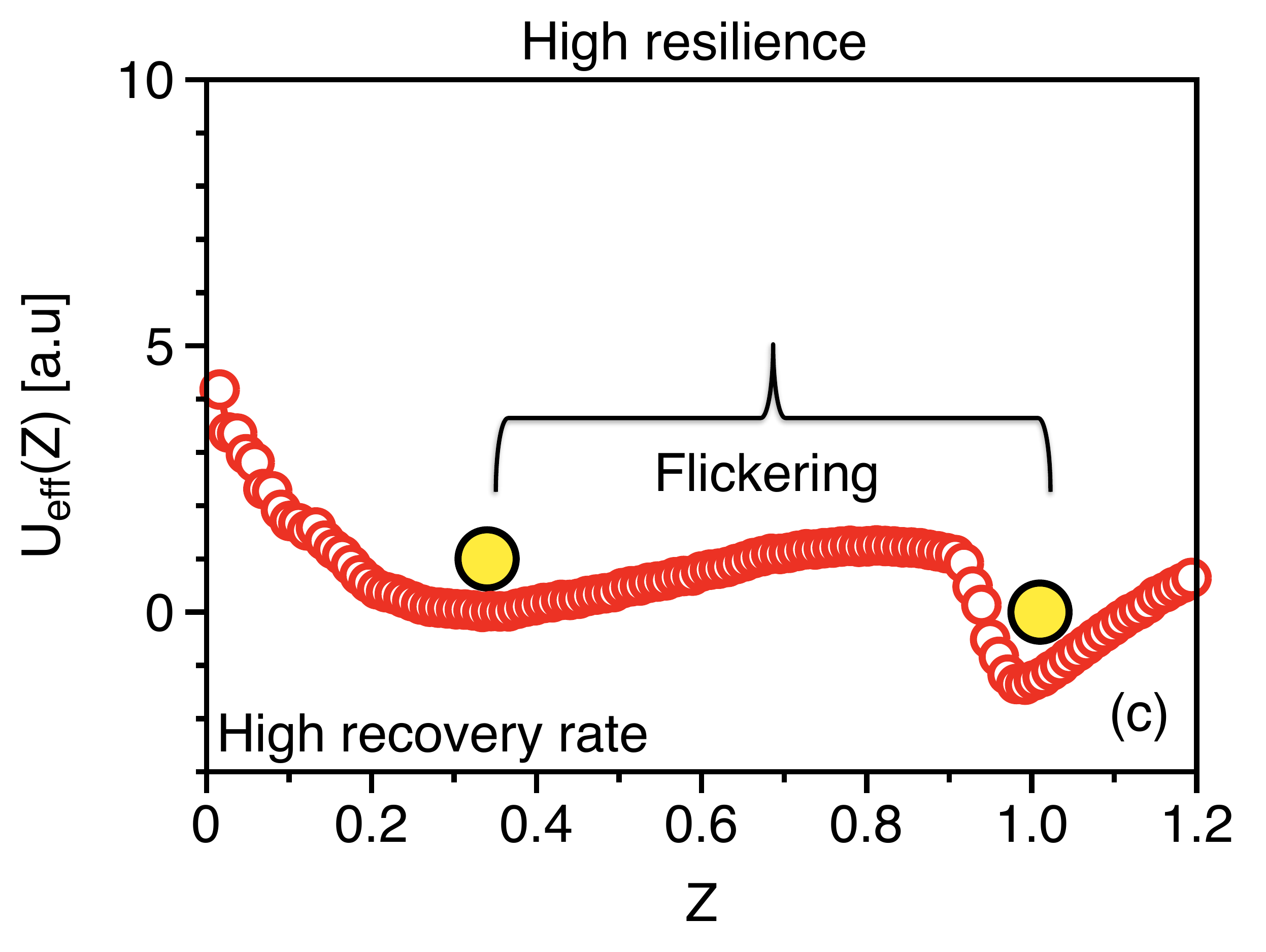}
\includegraphics[width=0.4\hsize]{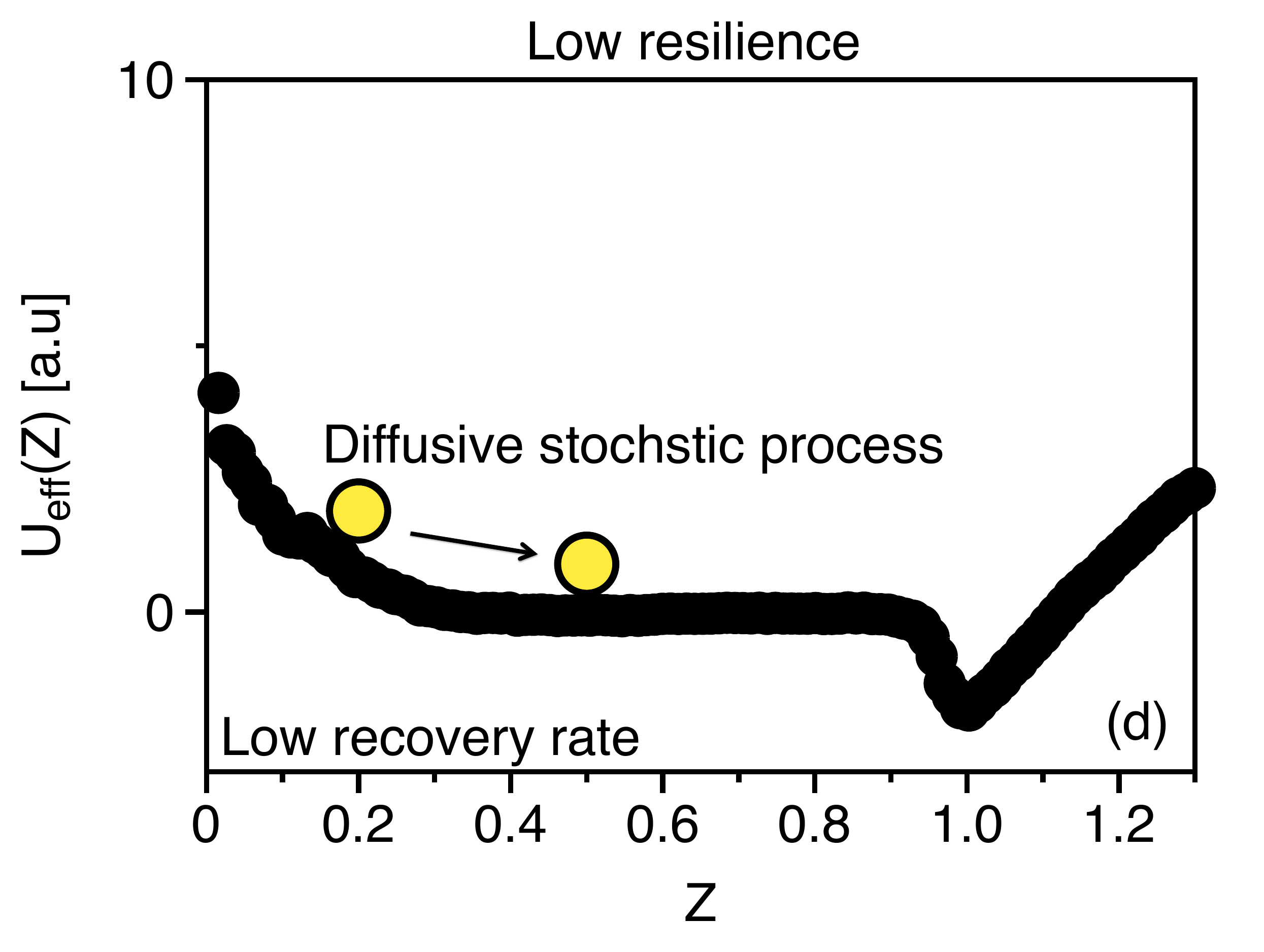}
\caption{The clear-sky index of (a) a single sensor and of (b) the solar field for Hawaii (S2). The single sensor time series has a flickering behaviour, while the average of the field exhibits a diffusive stochastic behaviour (without strong jumps). Illustration of the transition and critical slowing down when increasing the field size from (c) a single sensor to (d) the entire field. 
}\label{10}
\end{figure}

Beside the investigation of increment PDFs, in this section we investigate the dynamics of the renewable wind and solar variations. We aim to find out which dynamical feature leads to the emergence of large increments and how this alters with the geographical size.  As shown in Figs. \ref{10}a and \ref{10}b, the time series for a single sensor has a flickering behaviour, while for the field, it has a diffusive stochastic behaviour (without strong jumps). From these illustrations, clear changes in the  flickering behaviour of the data sets become obvious.

To study whether the rapid output fluctuations are jumpy or diffusive (persistent), we construct the effective potentials of corresponding time series after the methods explained in \cite{13,18,19,20,21-1}. The probability density functions provide the shape of the effective potential of time series as,

\begin{equation}
Prob(P) \sim \exp (-U_{eff} (P)).
\end{equation}

In Figs. \ref{10}c and \ref{10}d we plot the effective potential  $U_{eff} (Z)$  corresponding to the time series of Figs. \ref{10}a and \ref{10}b. The effective potential for a single sensor is asymmetric with a double-well structure. Note that the valleys in the effective potential represent stable attractors which are separated by a transition point (local maximum) for the single sensor at  $Z=0.8$ for solar data set (S2).  This double-well structure vanishes for the solar field data. 

The first minimum in the effective potential of Fig. \ref{10}c corresponds to a "cloudy" state, while the second minimum is related to a "clear sky" or "sunny" state. The depth of the minima correspond to the occupation probability, the deeper a minimum the higher the probability of this state. In Figs. \ref{10}c and \ref{10}d, it is shown that the increase of the number of sensors (the size of the solar field) leads to shallower potentials, and the barrier between the two minima approaches zero, causing a slowing down in the dynamics. For the solar irradiance data in Hawaii the behavioural transition occurs for a critical field size of about $ \sim  1 \times 1$~km$^2 $. As a consequence of this slowing down, the system has a longer memory and its dynamics are characterised by a small jump rate and a higher correlation time scale, as will be discussed next. 

\begin{figure}
\centering
\includegraphics[width=0.4\hsize]{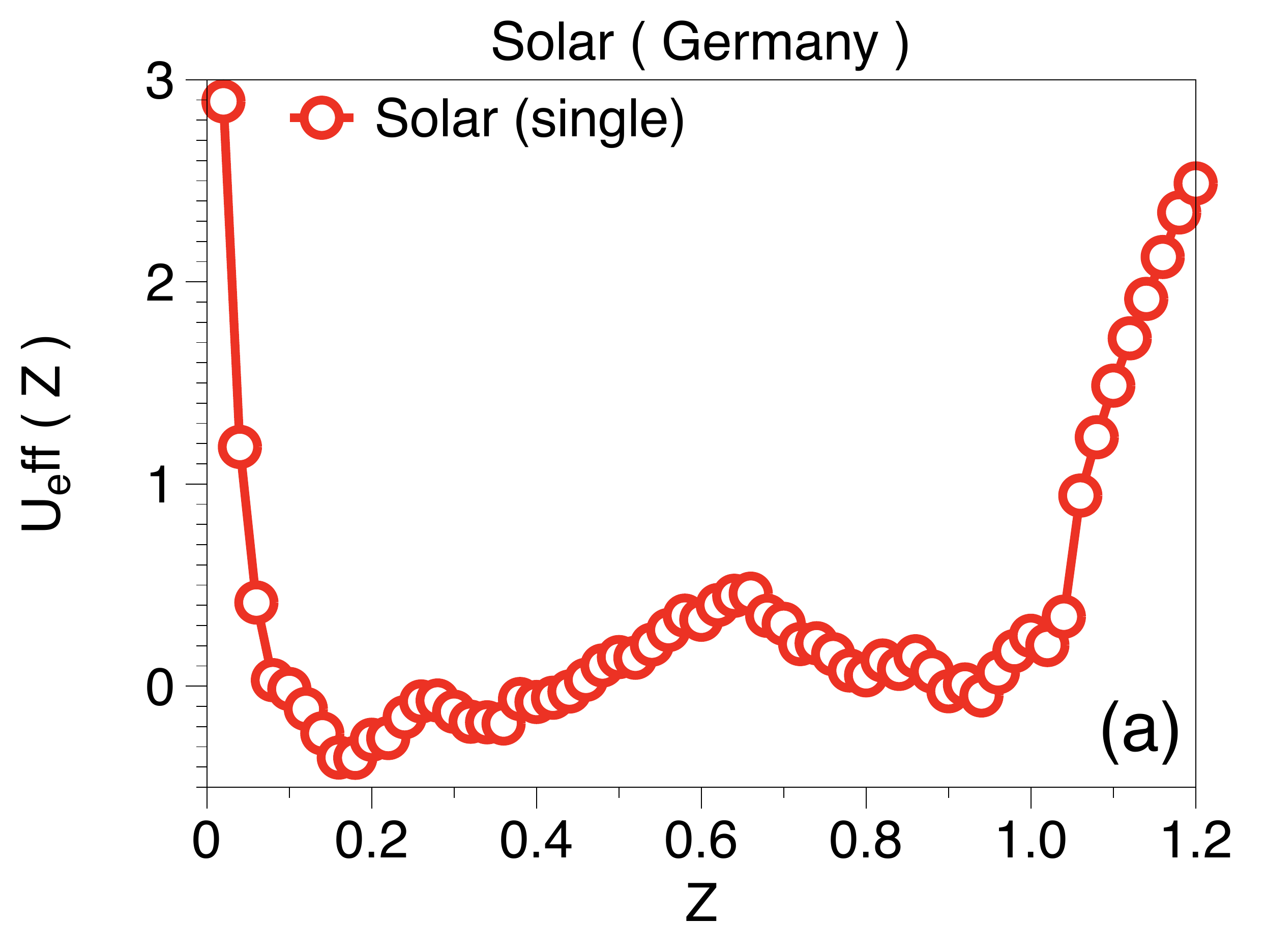}
\includegraphics[width=0.4\hsize]{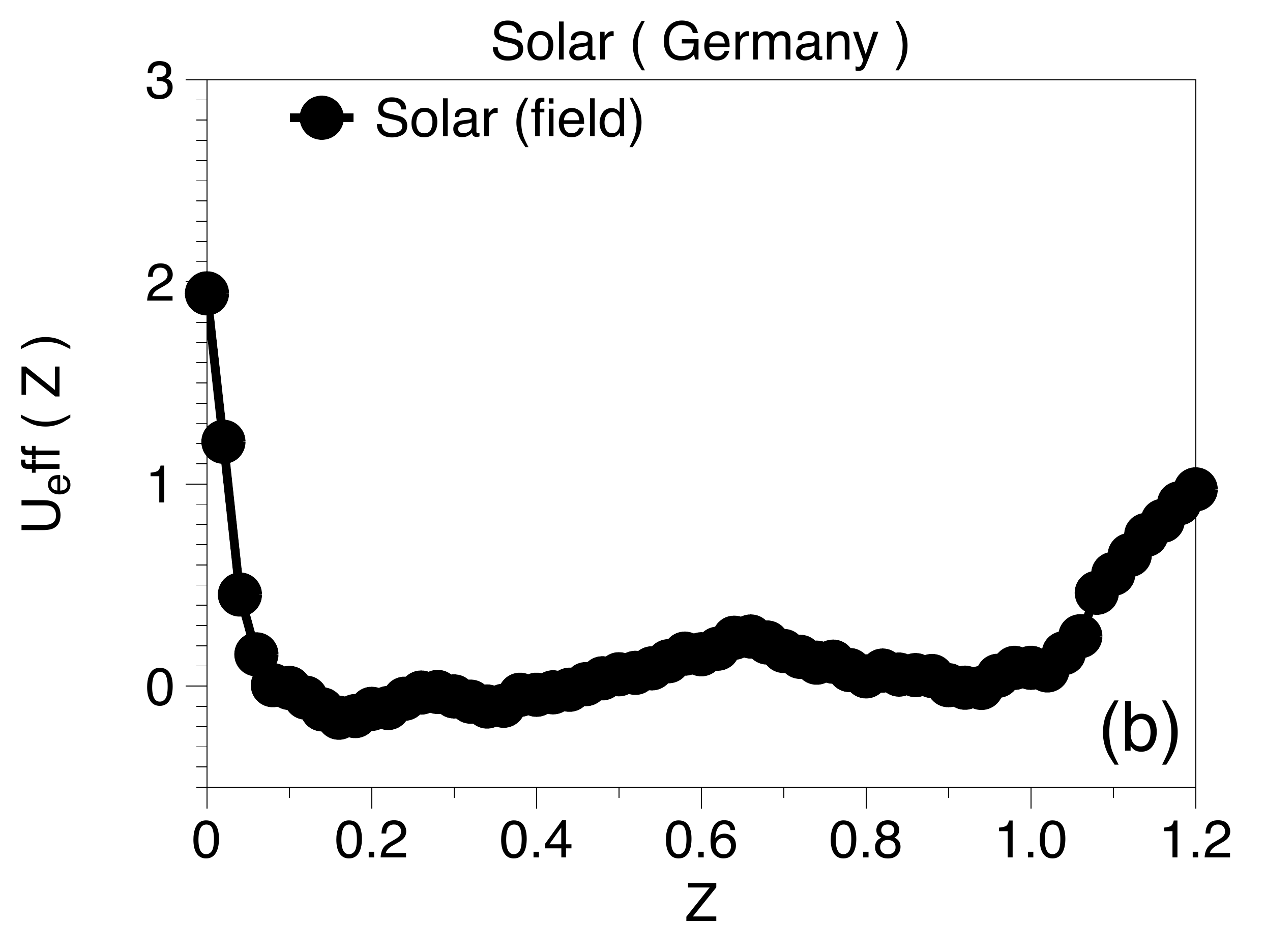}
\caption{Illustration of the transition and critical slowing down when increasing the field size from (a) a single sensor to the (b) entire field, Germany data set (S1).}\label{11}
\end{figure}

A similar trend exists for the data from the German solar field and with the transition point at $Z=0.65$ for the single sensor, as shown in Fig. \ref{11}.
However, in this case the field size is not large enough to detect the transition. This means that the critical field size is not a universal length scale and depends on the weather conditions of the area under investigation. The important observation is that larger fields have smoother clear-sky index fluctuations. A rapid change of dynamics with rapid ramp events remains for small field sizes. These results are interesting additional aspects to the changes in the intermittent behaviour of the power increment statistics as discussed in previous section, where we did not see an indication of such a clear change in the structure of the dynamics.

\begin{figure}
\centering
\includegraphics[width=0.5\hsize]{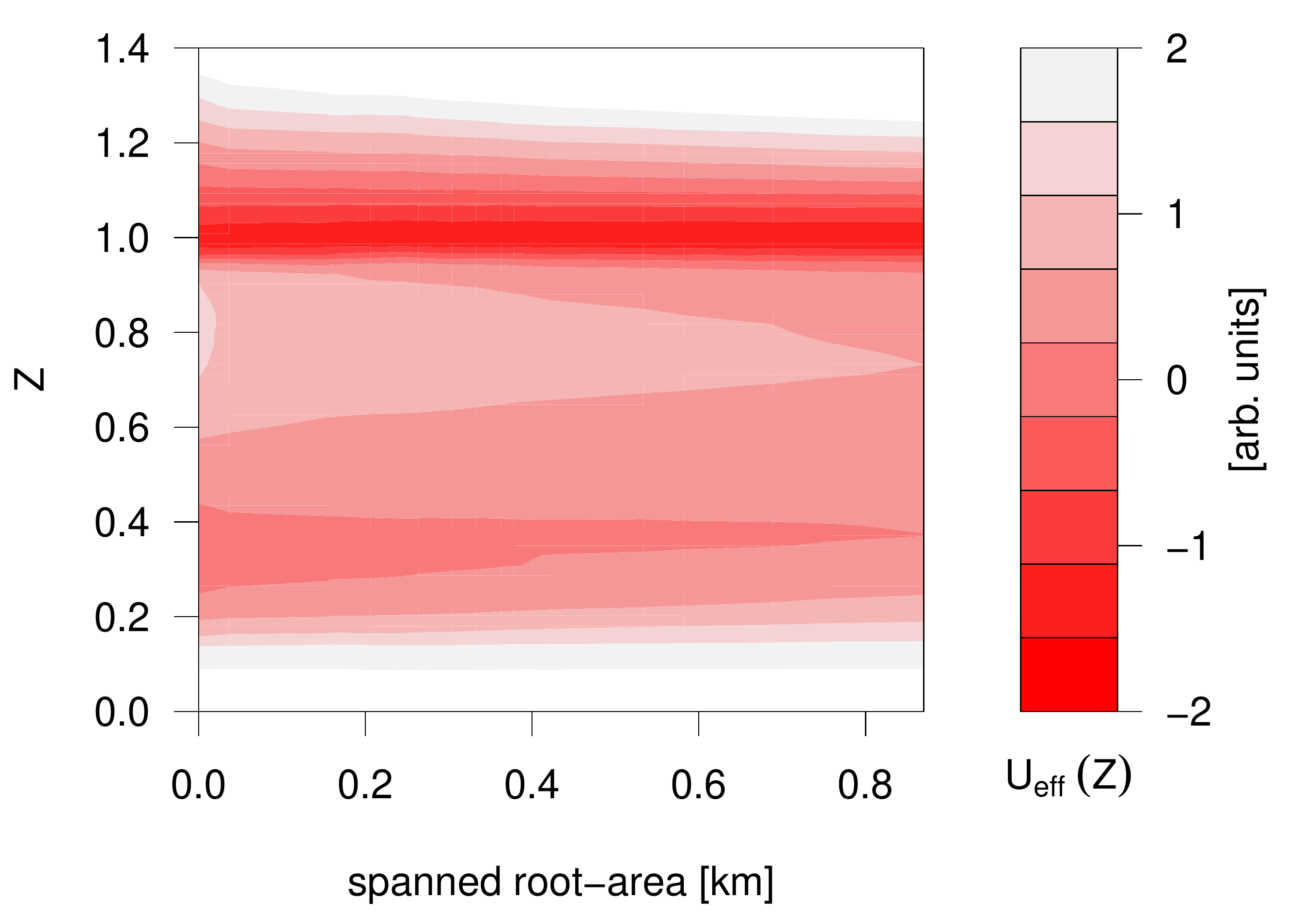}
\caption{A two-dimensional contour plot of the effective potential $U_{eff}(Z)$ of clear-sky index is plotted as a function of the field size. The data for this plot were measured in Hawaii.}\label{12}
\end{figure}

\begin{figure}
\centering
\includegraphics[width=0.4\hsize]{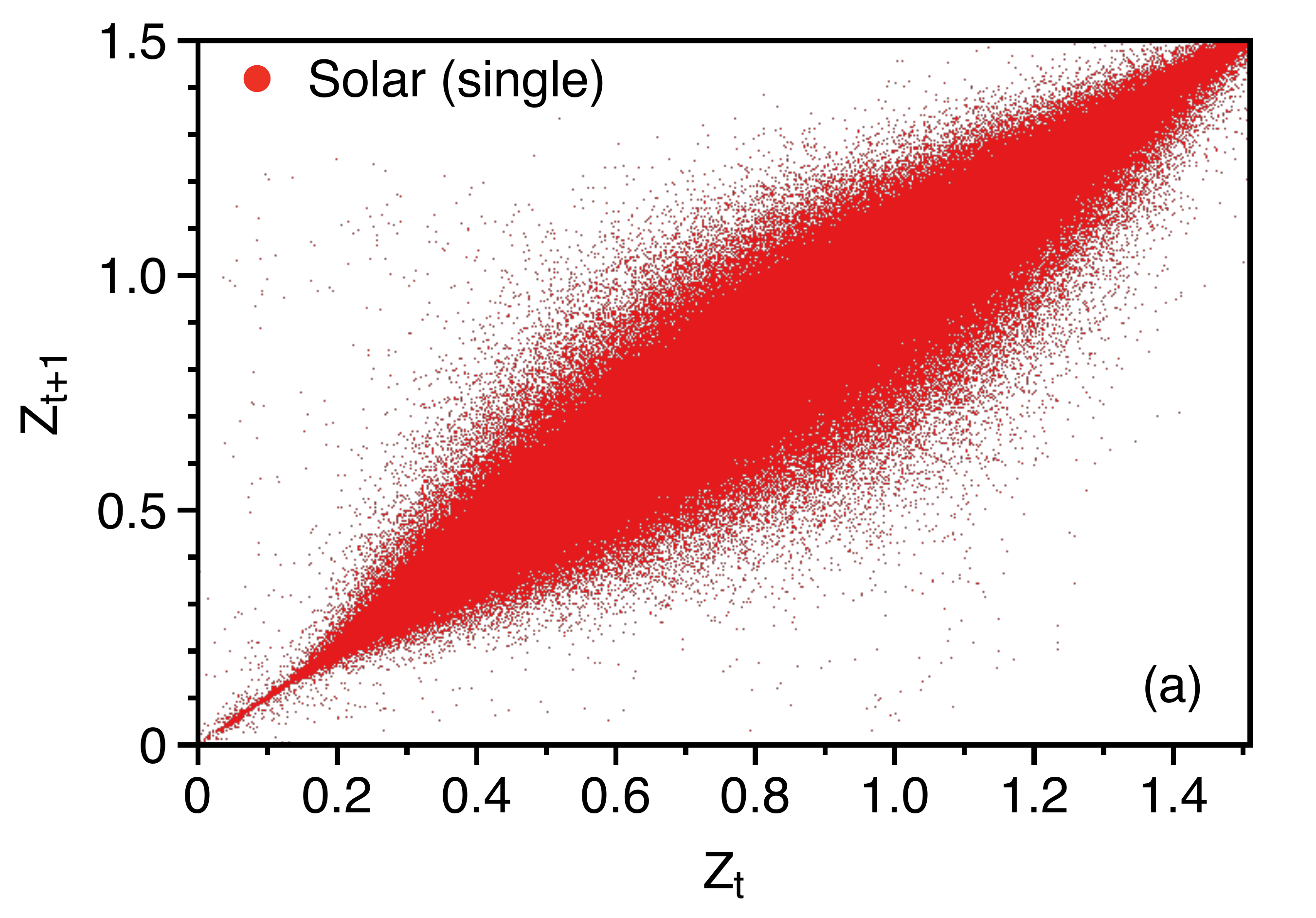}
\includegraphics[width=0.4\hsize]{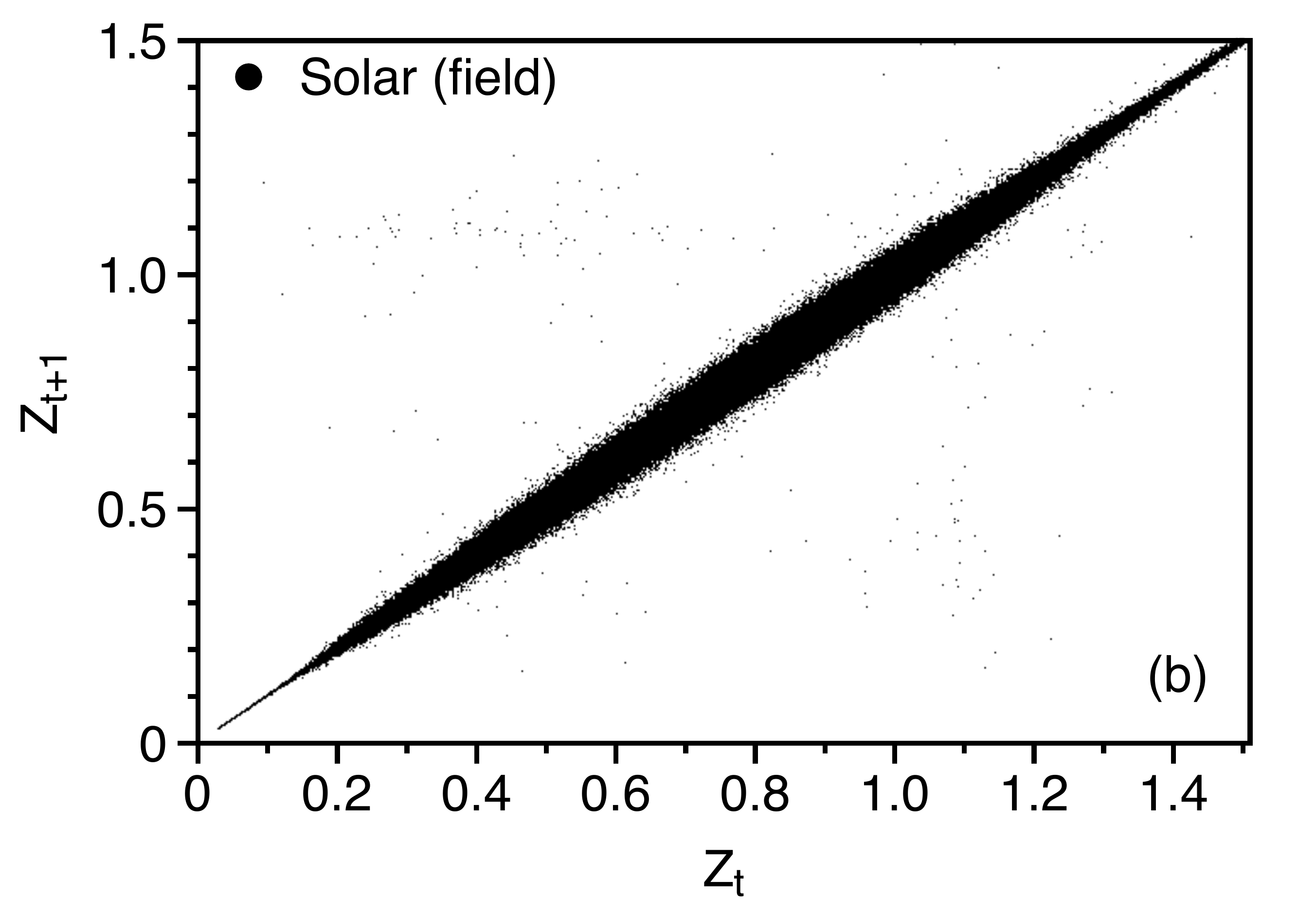}
\caption{The resulting dynamics for single sensor (a) and entire field (b) is characterised by a moderate correlation between the clear sky index at two subsequent times.}\label{21}
\end{figure}

In Fig. \ref{12}, a two-dimensional contour plot of the effective potential $U_{eff} ( Z )$ is plotted for various field sizes (estimated as the square root of the field area). It shows how the potential flattens as the spanned area increases, for clear-sky index $Z< 1$. Figs. \ref{21}a and \ref{21}b show the correlation between the clear sky index at two subsequent times ( $t$ and $t+1$ secs) for single sensor and solar field, respectively. For the entire field the resulting dynamics are characterised by a stronger correlation between subsequent states.

\begin{figure}
\centering
\includegraphics[width=0.5\hsize]{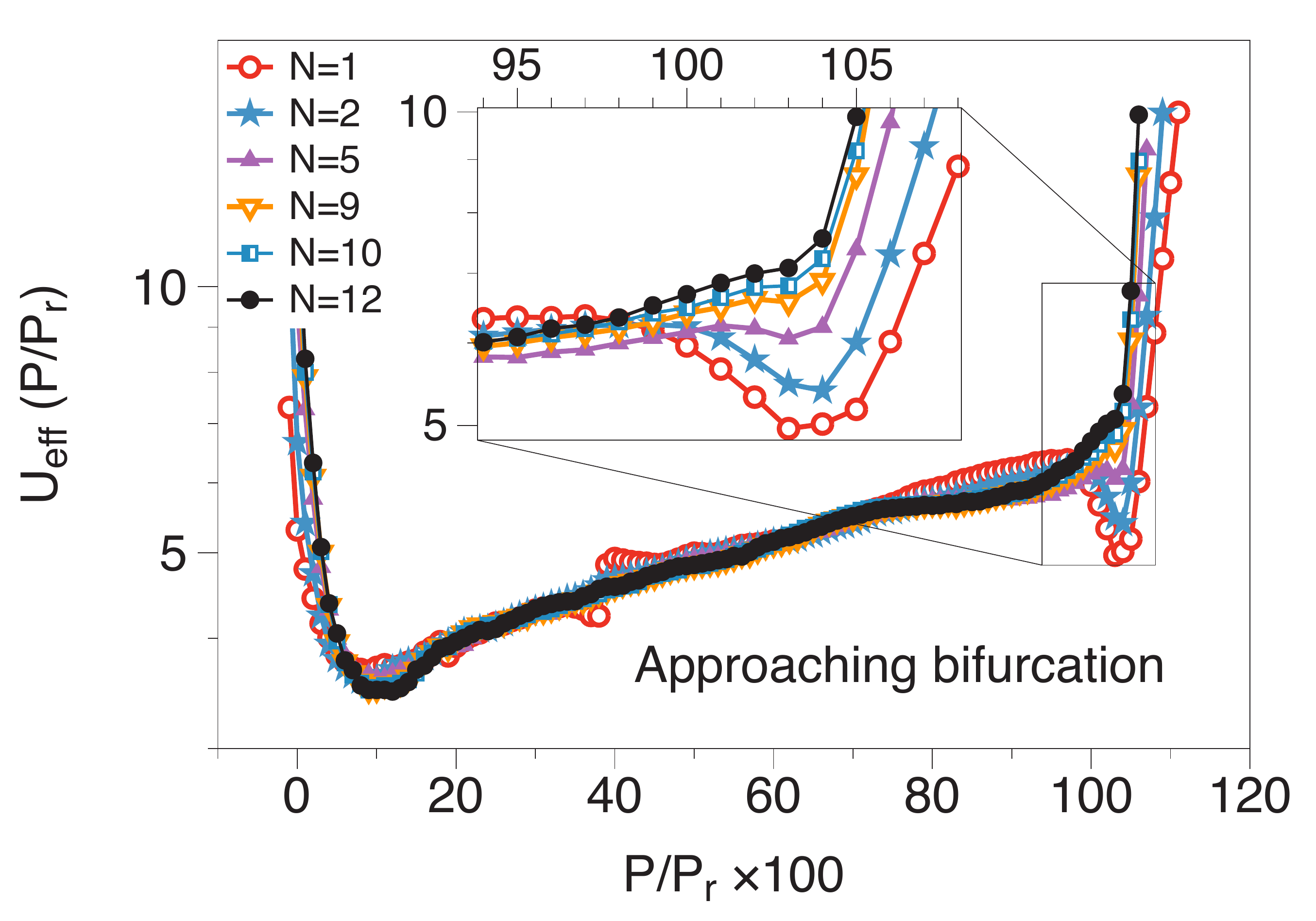}
\caption{Effective potential of the wind park $U_{eff} ( P/P_r )$ and its dependence on the number of wind turbines.}\label{13}
\end{figure}

In a similar way, in Fig. \ref{13}, we plotted the $U_{eff} (P)$ for a wind farm with a varying number of wind turbines and identify a similar transition as in the solar field. The distinct potential wells again represent two stable attractors, at about 10 \%  and 103 \% of the rated power for the single wind turbine. When increasing the number of wind turbines in the farm, the double-well structure changes to a potential with a single minimum at $\sim 10 \%$. The critical number of turbines for the behavioural transition is about $n_c \simeq 10$ turbines (with an area $\sim$ 4~km$^2$).  

In summary, based on the temporal analysis we found the interesting new aspect of the power dynamics changing from a bi-stable jumpy behaviour to a more diffusive one. As an important conclusion, increasing the field size solely suppresses the jumpy behaviour in the aggregated power output, but the non-Gaussian distributions of ramp events in terms of increment statistics remain even for country-wide installations.
 
\section{Suppressing the non-Gaussian statistics of wind and solar power} 

According to the results of the previous sections, both wind power and solar irradiance are characterised by abnormal statistics. 
Particularly on short time scales there are extreme power and irradiance fluctuations with high probabilities. 
Based on the temporal dynamics, accumulated renewable sources over smaller regions are more jumpy. 
Although, these multi-stable jumpy dynamics can be altered by combining more power units, the non-Gaussian character of renewable energies does not change in principle. 
Thus, building a reliable power supply in the presence of increasing shares of renewable energies remains as a challenge. 
In the actual discussion it is commonly accepted that technical solutions, such as fast reserves or storage systems in power supply are needed to overcome the intermittent fluctuations. 
In addition, intelligent technical solutions are promising as they may contribute directly to reduce the cost of energy (CoE) possibilities. 
These intelligent solutions are of high interest in the context of "smart grid" discussions. 
Based on the above presented insight, in the following we will present an idea of a simple modification of the dynamics, which enable us to decrease the intermittency of renewable sources in the range of seconds. 

We propose here, a time-delayed feedback method as an algorithm to generate the new power data sets based on the original data. This method is originated from the idea of storing a fraction $\alpha$ of power for a short while, and releasing it after a certain delay lag $T$. For this purpose, for instance we can assume that $N$ number of multiple wind or solar power plants are each equipped with suitable short-term storage and their aggregated power output equal to $P^*(t)=N^{-1}\sum_{i=1}^{N} p_{i}(t)$. In this way, the power output of the i-th renewable source $p_{i}(t)$ could change to

\begin{equation}
p_i (t ) ^{new} = (1-\alpha) p_i  (t)+\epsilon  P^*(t-T)
\end{equation}

where, in general, $\epsilon \leq \alpha$ ($\epsilon = \alpha$ for a power conserving model). Now, we analyse these new data sets to consider how much the intermittency of wind and solar power decreases in short time scales. 

\begin{figure}
\centering
\includegraphics[width=0.8\hsize]{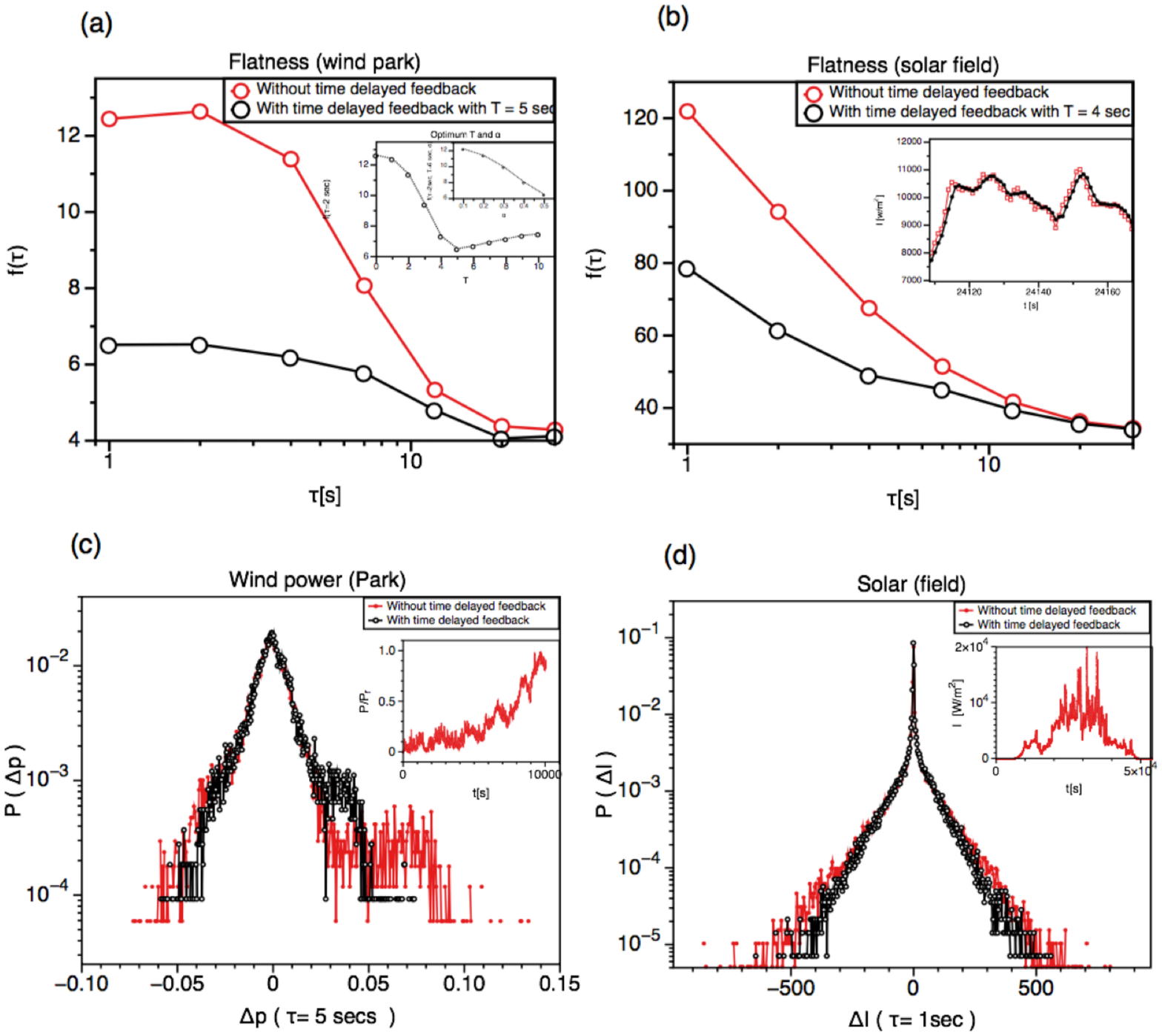}
 \caption{The results of the time-delayed feedback method to suppress the short term extreme events of a wind farm and solar field. Panels (a-d) show characteristic changes in stochastic dynamics of wind farm and solar field, their flatness (a,b) and probability distribution function of increments (c,d), when applying the time delayed feedback method to control the short time extreme events. The suppressing of extreme events is evident in all panels.
In the inset of panel (a) the optimum values of delay lag $T$ and amplification coefficient $\alpha$ in time delayed feedback method with minimising the flatness in time lag=2 seconds, is shown for a wind farm of 12 turbines. The optimum delay lag is $T=5$ with $\alpha=0.5$.
 In the inset of (b) the power output of the solar field is demonstrated, showing smoother dynamics when applying the time delayed feedback method. The results presented in panels (a,c) are derived from 10000~s of data with sample rate 1~Hz, belongs to a time interval during which the wind farm had strongly intermittent fluctuations, as shown in the inset (c). The solar data in panel (b) belongs to a very variable cloudy day in Hawaii (03.03.2011), see inset (d).}\label{30}
\end{figure}

The new cumulative power output $\sum_i ^N p_i (t ) ^{new}$ depends on the delay lag $T$ and saving factor $\alpha$. Their optimal values can be determined from minimisation of, for example, increment flatness. As an example, for W1 and S2 data sets we found that the optimal time delay-lag ranges between 2 and 5 seconds. For these $T$ values, the flatness of the short-term increment PDFs decreases most strongly with increasing the $\alpha$. For instance, with $T=5$~s, the flatness of increments decreases from $12.6$ to $6.5$ for the wind farm (W1), as shown in Fig. \ref{30}a. Results for increments of the solar field are plotted in Fig. \ref{30}b and \ref{30}d. The suppressing of strong non-Gaussian statistics is evident in the tails of the distributions, i.e.\ the undesirable extreme events are strongly influenced by our time-delayed feedback method.

For a possible application we suggest to use this time-delayed feedback method
as a control algorithm. Such a new control system could be based on electricity storage subsystems
like batteries or the rotational inertia of the rotor of wind turbines. It is known that batteries can age rapidly in this way (for further details see Ref. \cite{6-3} and references therein) and other technical problems may emerge.
We will leave a detailed technical discussion, corresponding realisations and method cost for the future.

\section{Concluding remarks}  

From a structural view point, power grids are complex networks which, due to economic factors, often run near their operational limits. 
The nature of renewable energies will add more and more fluctuations to this complex system, increasing intermittency and causing concern about the reliability and stability of the power supply. With the decreasing shares of conventional fossil and nuclear power systems, new concepts are needed in particular for short time aspects. In this work we have presented new statistical and dynamical details of wind and solar power fluctuations for the short time range of seconds to minutes which should be considered for designing the future power girds. 

The complexity of weather dynamics leads to short time non-Gaussian statistics in the power production from renewable sources.  There are different origins to observe the strong variability in wind and solar power fluctuations. Wind turbulence, which converts to wind power via wind turbine, is responsible for the short time scale intermittency of wind power output \cite{2}. For photovoltaics the dynamics of the clouds and their size distributions are the origin of its intermittent behaviour \cite{7-3}.  Most interestingly, the intermittency of nature will not be diminished by the transfer to power. For solar power one may argue that the shadows of the clouds cause an on-off threshold enhancing the fluctuations of the cloud structure, which is given by turbulence in the atmosphere. As a consequence of this increased complexity in the power dynamics, any central management of the grid is likely to become more and more difficult as the shares of renewable energies increase. Therefore the probability of having grid instabilities will increase, which may result in more frequent occurrences of extreme events like cascading failures resulting in large blackouts. Any strategy under discussion, like upgrading the existing power grid, the formation of virtual power plants combining different power sources, introducing new storage capacities and intelligent "smart grid" concepts, etc., will further increase the complexity of the existing systems and have to be based on the detailed knowledge of the dynamics of these renewable energies. Investigations of power grid stability in the presence of stochastic renewable sources, including their extreme events, provide a new emerging field of research which is a combination of these so far disconnected fields of work.

In this contribution we characterise the short time non-Gaussian statistics behaviour of wind and solar power, using the increment statistics and effective potential of dynamics. 
We find distinct behaviour of wind power and solar irradiance on different time scales, and quantify the likelihood of certain power fluctuations by parametrisation of increment PDFs. 
Furthermore, distinguishing jumpy and diffusive characteristics of short-term fluctuations may pave the way to the design and robust evaluation of power grid stability. The short time jumpy power output of small power units will demand more sophisticated methods to compensate for their on-off type behaviour and necessitates quick action in the order of seconds for solar, and a few minutes for wind power in response to observed power variability. Finally, we show that a simple dynamic variation using a time-delayed feedback method in the management of intermittent renewable sources will strongly suppress the non-Gaussian statistics. This method shows that the intermittent nature of renewable energies might not be a big problem if the intermittency is properly characterised. Otherwise it definitely might lead to grave grid problems. Because of the statistical approach presented in this article, we considered only the statistical changes in the time-delayed power and avoided technical discussions.

We propose our profound statistical analysis to be included in the guidelines of power systems to guarantee an optimal design of resilient power grids. The challenge will be to fine tune the intelligent management tools, as well as technological possibilities, to achieve a stable and low cost power system that can handle the intermittent renewable sources of power efficiently. 

{Acknowledgements:}
The Lower Saxony research network 'Smart Nord'
acknowledges the support of the Lower Saxony Ministry of Science and
Culture through the 'Nieders{\"a}chsisches Vorab' grant programme (grant ZN
2764/ZN 2896). We acknowledge also the National Renewable Energy Laboratory in the United States for providing the data in Hawaii. 
BSRN data was kindly made available by the World Radiation Monitoring 
Center (WRMC) and we particularly acknowledge Mohamed Mimouni and Xabier 
Olano, Managers of BSRN stations of Tamanrasset (Algeria) and Cener 
(Spain), respectively. We also acknowledge Deutsche Windtechnik AG Bremen for providing us with wind turbine data. We would like to thank, K. Aihara, L. von Bremen, J. Davoudi, O. Kamps,  H. Kantz, L. Kocarev, J. Kurths, P. Lind, N. Nafari, J. F. Pinton, S. Rahvar, A. Rostami, P. Rinn, M. Sahimi, K. Schmietendorf, M. Sonnenschein and K. R. Sreenivasan for important comments and discussions.

\newpage 



\end{document}